\documentclass[prb,reprint,superscriptaddress]{revtex4-1}
\usepackage[caption=false]{subfig}
\usepackage[version=4]{mhchem}
\usepackage{graphicx,color}
\usepackage[usenames,dvipsnames,svgnaes,table]{xcolor}
\usepackage{bm}
\usepackage{amsmath}
\usepackage{amssymb}
\usepackage{mathtools}
\usepackage{siunitx}
\usepackage{multirow}
\usepackage{array}
\usepackage{color}
\usepackage{longtable}
\usepackage{xfrac}
\usepackage{hyperref}
\hypersetup{colorlinks=true,
            linkcolor=red,
            citecolor=blue,
            urlcolor=orange}
\usepackage{graphicx, type1cm, lettrine}
\newcolumntype{K}[1]{>{\centering\arraybackslash}p{#1}}
\setcitestyle{numbers,round}

\begin{document}

\title{Ternary mixed-anion semiconductors with tunable band gaps from machine-learning and crystal structure prediction}

\author{Maximilian Amsler}
\email{amsler.max@gmail.com}
\affiliation{Laboratory of Atomic and Solid State Physics,
Cornell University, Ithaca, New York 14853, USA}

\author{Logan Ward}
\altaffiliation{Present address: Computation Institute, University of Chicago, Chicago, IL 60637, USA}
\affiliation{Department of Materials Science and Engineering, Northwestern
University, Evanston, Illinois 60208, USA}

\author{Vinay I. Hegde}
\affiliation{Department of Materials Science and Engineering, Northwestern
University, Evanston, Illinois 60208, USA}

\author{Maarten G. Goesten}
\affiliation{Department of Chemistry and Chemical Biology,
Cornell University, Ithaca, New York 14853, USA}

\author{Xia Yi}
\affiliation{Center for Nanoscale Materials, Argonne National Laboratory, Argonne, Illinois 60439, USA}

\author{Chris Wolverton}
\email{c-woverton@northwestern.edu}
\affiliation{Department of Materials Science and Engineering, Northwestern
University, Evanston, Illinois 60208, USA}

\date{\today}

\begin{abstract}

We report the computational investigation of a series of ternary \ce{X4Y2Z} and \ce{X5Y2Z2} compounds with X=\{Mg, Ca, Sr, Ba\}, Y=\{P, As, Sb, Bi\}, and Z=\{S, Se, Te\}. The compositions for these materials were predicted through a search guided by machine learning, while the structures were resolved using the minima hopping crystal structure prediction method. Based on \textit{ab initio} calculations, we predict that many of these compounds are thermodynamically stable. In particular, 21 of the \ce{X4Y2Z} compounds crystallize in a tetragonal structure with  \textit{I-42d} symmetry, and exhibit band gaps in the range of 0.3 and 1.8~eV, well suited for various energy applications. We show that several candidate compounds (in particular  \ce{X4Y2Te} and \ce{X4Sb2Se}) exhibit good photo absorption in the visible range, while others (e.g.,  \ce{Ba4Sb2Se})  show excellent thermoelectric performance due to a high power factor and extremely low lattice thermal conductivities. 
\end{abstract}
\maketitle

\section{Introduction}
Computational approaches are being employed at an increasing rate to discover and design novel materials with tailored properties to tackle global environmental challenges. Traditionally, two approaches are frequently employed for these efforts: while high-throughput density functional theory (DFT) calculations~\cite{saal_materials_2013, kirklin_oqmd_2015, Jain2013,curtarolo_aflowlib.org_2012, curtarolo_high-throughput_2013, toher_aflow_2017} are popular to explore the chemical space, crystal structure prediction schemes (CSP)~\cite{oganov_modern_2010} are used to determine the ground state configuration at a given composition by globally minimizing the (free) energy.

Very recently, novel methods based on materials informatics and machine learning (ML) models have emerged to assist the search for materials with improved properties in industrially relevant applications. Trained on available materials data either from experimental observations or DFT datasets (like the OQMD~\cite{saal_materials_2013, kirklin_oqmd_2015}, Materials
Project~\cite{Jain2013}, and AFLOWlib~\cite{curtarolo_aflowlib.org_2012}), most of these ML models aim at predicting promising chemical subspaces for materials with favorable properties~\cite{ward_atomistic_2017}. Examples include models for melting temperatures~\cite{seko_machine_2014}, materials thermodynamics~\cite{ghiringhelli_big_2015,meredig_combinatorial_2014,deml_predicting_2016,bartel_gibbsML_2018,isayev_fragment_2017,faber_machine_2016,seko_matrix-_2018,schmidt_thermo_2017},  mechanical properties of alloy systems~\cite{bhadeshia_performance_2009,chatterjee__2007,xue_sma_2016}, superconductivity~\cite{stanev_machine_2018}, formation of metallic glasses~\cite{ward_machine_2018}, and  electronic properties of semiconductors and insulating materials~\cite{dey_informatics-aided_2014,pilania_machine_2016,ward_general-purpose_2016,huan_motif_2015,sparks_te_2015}.

Despite these promising developments, a major drawback of such ML schemes is that they often predict promising chemical compositions without providing any information on the underlying crystal structure. However, the knowledge of the atomic arrangement in a crystal lattice is crucial for any further computational assessment from first principles calculations. Hence, it is desirable to augment any ML prediction of composition alone in some way with a determination of the corresponding ground state crystal structure to (a) validate the ML model and (b) to facilitate further computational investigations in terms of materials properties.

To tackle this issue, we combine recently developed ML models by Meredig~\textit{et al.}~\cite{meredig_combinatorial_2014} and Ward~\textit{et al.}~\cite{ward_general-purpose_2016} to predict the formation enthalpies and band gap energies with a sophisticated CSP scheme, the minima hopping method (MHM)~\cite{goedecker_minima_2004,amsler_crystal_2010}. The MHM takes as the input the chemical composition from the ML model and optimizes the potential energy, providing promising candidates for the ground state structure. We apply this approach on a subset of complex, ternary chemistries that are predicted by the ML to exhibit finite band gaps in a particularly promising range for many energy applications, including photovoltaics, photocatalysis, and thermoelectrics.  We predict a class of ternary compounds with \ce{X4Y2Z} composition that are both thermodynamically stable and have band gaps between 0.3 and 1.8~eV. Our calculations show that several compounds are indeed promising for photovoltaic applications, most notably  \ce{X4Y2Te} and \ce{X4Sb2Se}, with a strong overlap of the absorption and solar spectrum. At the same time, several candidates exhibit large power factors and low thermal conductivities due to the underlying structural complexity, rendering them promising candidates as thermoelectric materials. In fact, \ce{Ba4Sb2Se} shows an extremely low thermal conductivity of $\kappa=0.61$~Wm$^{-1}$K$^{-1}$ at 300~K, close to the best thermoelectric material known to date, SnSe~\cite{SnSe,doi:10.1038/nphys3492}.

\section{Methods}

\subsection{Machine learning model}

Machine learning models are composed of a set of training data, a representation that converts the training data into a form suitable for machine learning (i.e., tensors), and a machine learning algorithm. For this work, we use the training set, representation, and algorithms described in Ref.~\onlinecite{ward_general-purpose_2016}. Specifically, we use the formation enthalpy and band gap energy data from the Open Quantum Materials Database (OQMD)~\cite{saal_materials_2013, kirklin_oqmd_2015} to train our models. From the entire OQMD, we only use the lowest energy structure at each composition - a total of 228,649 entries. Our representation is composed of 145 different attributes of the composition of each entry, such as the mean melting temperature of the constituent elements and whether the compound is charge-balanced. We use a hierarchical machine learning model based on decision trees that was found in Ref. ~\cite{ward_general-purpose_2016} to predict the band gap energies, and the Random Forest algorithm~\cite{breiman_randomforest_2001} to predict formation enthalpies. All models are created using the Material Agnostic Platform for Informatics and Exploration (Magpie)~\cite{ward_general-purpose_2016}, and are available as Supplementary Information to this paper and on GitHub.~\cite{mhm_github}

\subsection{Structural searches}
We employ the Minima Hopping Method (MHM)~\cite{goedecker_minima_2004,amsler_crystal_2010} to explore candidate structures for the most promising compositions proposed by the ML model. The MHM implements a reliable algorithm to explore the low-lying portions of a potential energy surface solely given the chemical composition. Consecutive, short molecular dynamics (MD) escape trials are employed to overcome energy barriers, followed by local geometry optimizations. The Bell-Evans-Polanyi principle is exploited by aligning the initial MD velocities along soft-mode directions in order to accelerate the search~\cite{roy_bell-evans-polanyi_2008,sicher_efficient_2011}. In the past, the MHM has been successfully employed to predict or resolve the structure of a wide class of materials~\cite{amsler_crystal_2012,flores-livas_high-pressure_2012,amsler_novel_2012,huan_thermodynamic_2013,clarke_discovery_2016,clarke_creating_2017,amsler_prediction_2017,amsler_exploring_2018}.

\subsection{Formation enthalpies and electronic struture}
The training data for the ML model is generated from DFT in a high-throughput fashion and is freely available in the OQMD. All DFT calculations are performed with the Vienna Ab initio Simulation Package (VASP)~\cite{kresse1993ab, kresse1996efficiency,kresse1996efficient} within the projector augmented wave (PAW) formalism~\cite{blochl1994projector, kresse_paw_1999} in conjunction with the PBE parameterization of the generalized gradient approximation to the exchange correlation functional~\cite{perdew1996generalized}. We use $\Gamma$-centered $k$-point meshes with $\approx 8000$~$k$-points per reciprocal atom and a plane-wave cutoff energy of 520~eV. All structural relaxations are carried out by taking into account the atomic and cell degrees of freedom until the force components on the atoms were within 0.01~eV/\AA{}, and stresses were within a few kbar. Similarly, the phases resulting from the structural searches are refined with the same DFT settings to obtain comparable formation energies on the same technical footing.

The band gaps used to train the ML model are based on the PBE functional. Although semi-local functionals are known to underestimate the band gaps, the overall correlation with respect to different materials chemistries is well reproduced, i.e., the error is systematic and corresponds to a constant shift of the band gap values.

To analyze the chemical bonding, we study the Crystal Orbital Overlap Population (COOP), which decomposes the electronic density of states into bonding and anti-bonding contributions. For this purpose, we use the the ADF Band package~\cite{te_velde_precise_1991,wiesenekker_quadratic_1991,franchini_becke_2013,franchini_accurate_2014}, which employs basis sets of numerical atomic orbitals together with Slater-type orbitals. We use double zeta polarized ZORA basis functions (ZORA-DZP). This basis set is found to be sufficiently  converged by carefully inspecting the band structures. \textit{Ab initio} COOPs obtained from linear combination of atomic orbitals (LCAO) are especially sensitive to numerical issues arising from basis set over-completeness due to linear dependencies in construction of the overlaps.

\subsection{Optical properties}

The optical absorption properties for selected candidate materials are evaluated by computing the frequency dependent dielectric function $\epsilon(\omega)$, using the Heyd-Scuseria-Ernzerhof (HSE06) hybrid functional~\cite{paier_screened_2006,heyd_energy_2005,heyd_erratum:_2006,paier_erratum:_2006}. We employ a cutoff energy of 270~eV and a $6\times6\times6$ $k$-point mesh, and include at least 124 virtual bands (corresponding to more than twice the number of occupied states).

\subsection{Electronic transport}

The electronic transport properties are computed by solving the Boltzmann transport equation within the constant relaxation time approximation using the Boltztrap code~\cite{Madsen200667}. This approximation is commonly applied for doped semiconductors and assumes that the electronic relaxation time varies little with energy on the scale of $k_BT$~\cite{singh_doping-dependent_2010}. The PBE band structures are resolved on a $35\times35\times35$ $k$-point mesh. Typical values for the relaxation time is around $\tau=10^{-14}$~s~\cite{chen_understanding_2016}. Here, we use a constant relaxation time of $\tau=3.4\times 10^{-14}$~s in accordance with the work of Bilc~\textit{et al.}~\cite{bilc_low-dimensional_2015}.

\subsection{Lattice dynamics}
Phonons are computed by using the finite difference approach to obtain the second order interatomic force constants. Supercells of dimension $2\times 2\times 2$ are used, and all calculations are performed with the \texttt{Phonopy} package~\cite{phonopy}.

The thermal transport calculations are carried out by taking into account three-phonon interactions. We use the recently developed compressive sensing lattice dynamics~\cite{PhysRevLett.113.185501} (CSLD) technique to obtain the third order force constants (FC). For every compound, the FC are fitted to atomic forces from 40 randomly displaced supercells containing 756 atoms, using a cutoff energy of 400~eV and a $\Gamma$-only $k$-points setting. Cutoff radii of 14~\AA~and 5~\AA~are used to truncate the 2- and 3-body interactions, respectively. The FC are subsequently fed into the ShengBTE package~\cite{ShengBTE_2014} to iteratively solve the linearized Boltzmann phonon transport equation.

\section{Results and Discussion}
\subsection{Discovery of New Ternary Semiconductors}

\begin{figure}[htb!]
\centering
\includegraphics[width=0.8\columnwidth]{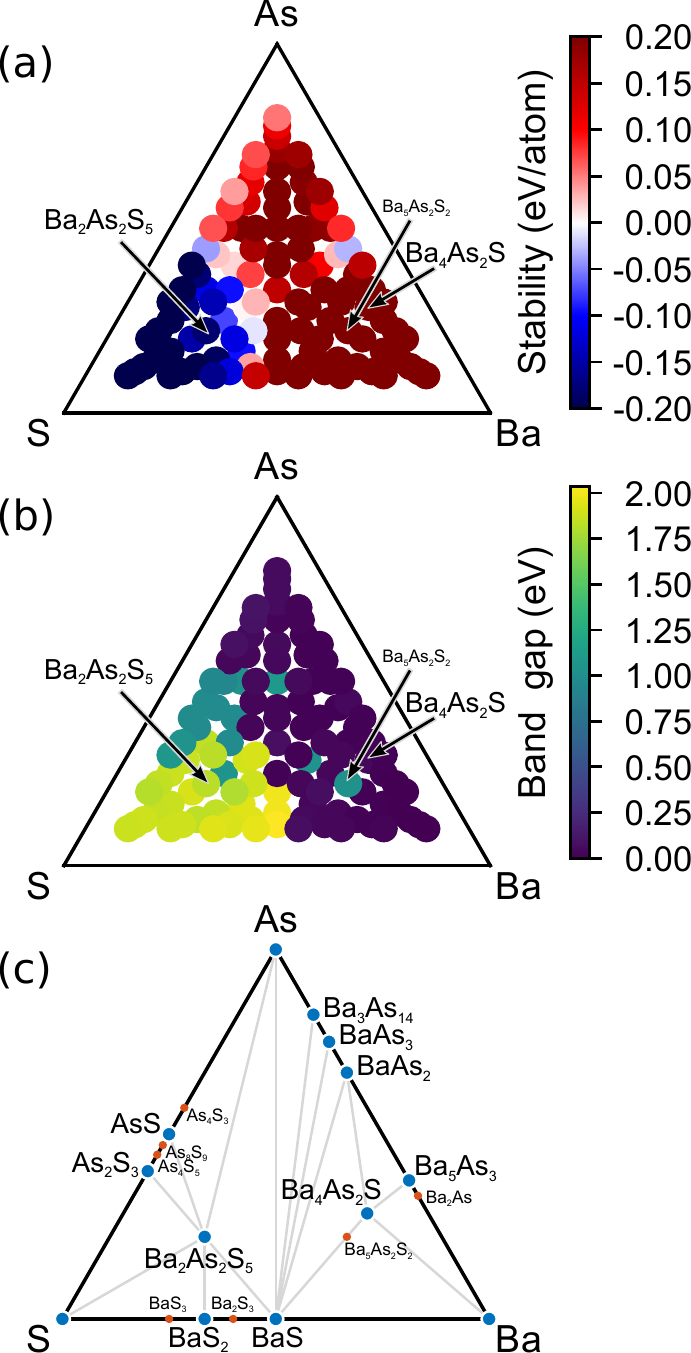}
\caption{(a) Stability and (b) band gap energy of various compositions in the Ba--As--S ternary system predicted using ML models.
The stability is determined by first predicting the formation enthalpy of a composition using ML, and then measuring the hull distance obtained from the OQMD. 
We find two regions with negative formation enthalpies, which indicates the presence of a potential compound currently missing from the convex hull of the OQMD.
The band gap energy model predicts that a significant fraction of the Ba--As--S compounds have band gap energies of approximately 1~eV. Panel (c) shows the convex hull  construction based on DFT energies, including all experimentally observed phases and the putative ground states from MHM simulations for \ce{Ba4As2S} and  \ce{Ba5As2S2}. Blue and orange circles denote thermodynamically stable and unstable phases, respectively.
}\label{fig:MLTernary}
\end{figure}

Our initial search for new semiconducting materials is guided by the compounds predicted to be stable by the ML model of Meredig~\textit{et al.}~\cite{meredig_combinatorial_2014}. One of the compounds predicted to be stable is \ce{Ba2As2S5}. As shown in Fig.~\ref{fig:MLTernary}(a), we are able to confirm this earlier prediction using the newer ML models from Ward~\textit{et al.}~\cite{ward_general-purpose_2016}. We then evaluate all Ba-As-S compositions with less than 12 atoms per formula unit to detect other potentially-stable materials in this system. We find that there exists a second region of stable compounds in the proximity of the Ba--As binary system. We then evaluate the Ba--As--S system with the band gap energy ML model from Ward~\textit{et al.}~\cite{ward_general-purpose_2016} and find that a large fraction of this ternary system is predicted to have band gaps of approximately 1~eV (see Fig.~\ref{fig:MLTernary}(b)). Both the stability and band gap energy predictions indicate that the Ba--As--S system contains promising, yet-undiscovered semiconducting materials.

Given the two composition regions of interest, we then explore the literature to further refine a list of candidate materials. 
We find that \ce{Ba2As2S5} is an already known material that was however not present in the OQMD at the time our training set was assembled. This result provides strong confidence that our ML models indeed yield reliable predictions.
On the other hand, to the best of our knowledge, there exist no experimental evidence of a ternary compound near the Ba-- and As--rich portion of the phase diagram.
To generate a first list of candidates, we enumerate compositions that correspond to the common oxidation states of \ce{Ba^2+}, \ce{As^3-} and \ce{S^2-}. 
Starting with the smallest possible formula units (f.u.) to render CSP computationally tractable, we are left with the ternary compositions \ce{Ba4As2S} and \ce{Ba5As2S2}, with number of atoms per f.u.  of 7 and 9, respectively.

\begin{figure}[htb!]
	\centering
	\includegraphics[width=1\columnwidth]{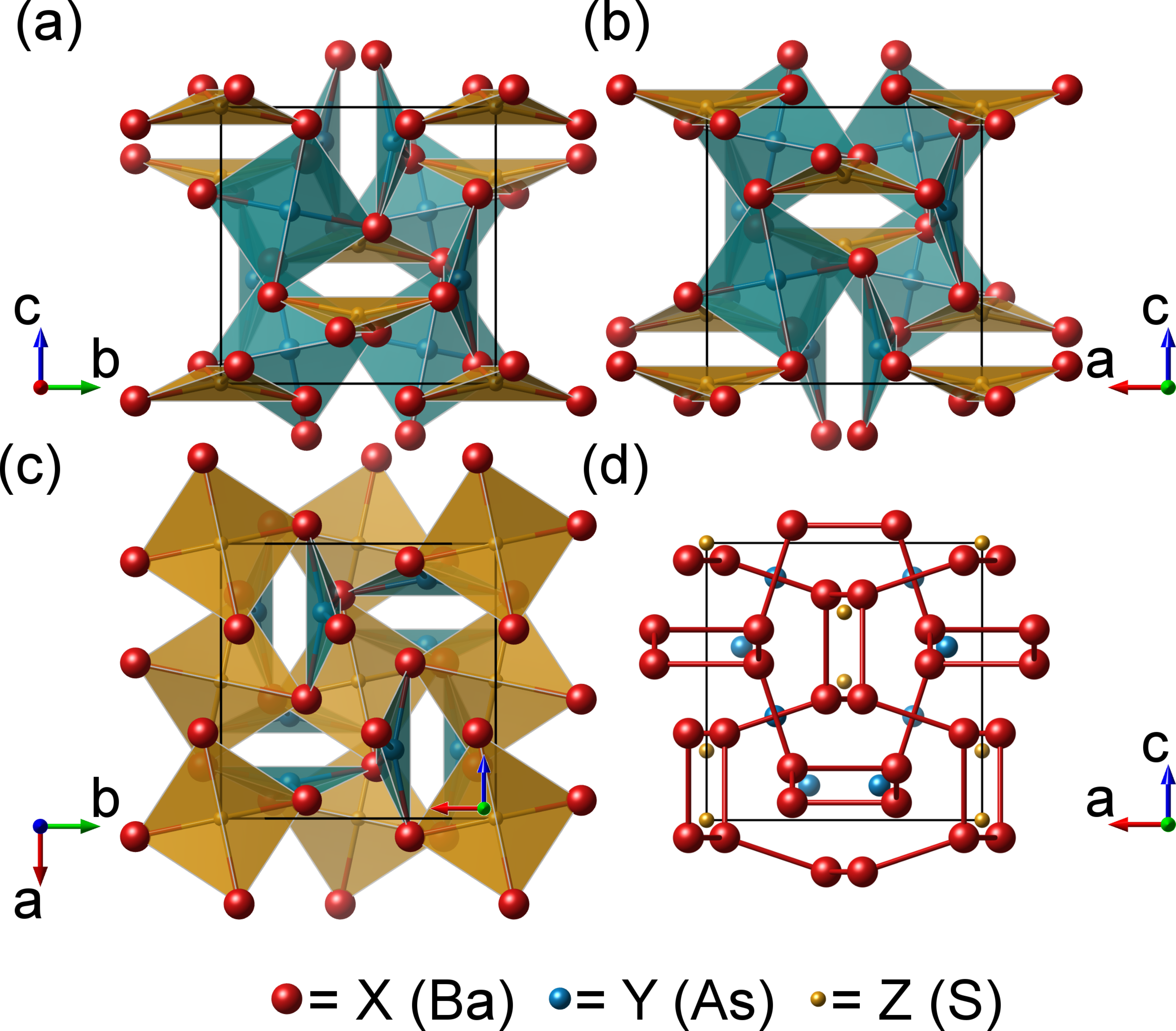}
	\caption{The structural details of the \ce{X4Y2Z} compounds with \textit{I-42d} symmetry, predicted through the MHM simulations. Here, \ce{Ba4As2S} serves as a representative compound. Panels (a), (b) and (c) show the crystal structure from three different perspectives with the distorted, corner sharing tetrahedra of \ce{XY4} (\ce{BaAs4},  blue) and  \ce{XZ4}  (\ce{BaS4}, yellow) in a polyhedral representation. Panel (d)  shows the two enantiomeric networks formed by the X (Ba) atoms.}\label{fig:Structure}
\end{figure}

Based on this initial assessment, we perform MHM calculations with two f.u. for the two systems \ce{Ba4As2S} and \ce{Ba5As2S2}. In addition, we also explore the combination of all elemental substitutions in \ce{X4Y2Z} and \ce{X5Y2Z2} with X=\{Mg, Ca, Sr, Ba\}, Y=\{P, As, Sb, Bi\}, and Z=\{O, S, Se, Te\}, i.e., 64 distinct compositions. Each of the MHM runs is terminated after finding some 100 distinct structures (corresponding to roughly 200 MHM iterations). For many systems of similar size, this number of iterations might not be sufficient to conclusively determine the ground state structure. Therefore,  in a second step, we adopt a high-throughput substitutional search to validate our predictions. For this purpose, we select the lowest energy candidates within each of the MHM runs in the 64 distinct compositions as prototype structures. These prototypes are then decorated with the elements of the remaining 63 compositions before performing a single, local geometry relaxation with refined parameters. Among those, the lowest energy structure \ce{X4Y2Z} and \ce{X5Y2Z2} are selected as the final candidate phases for the putative ground states.

To evaluate the thermodynamic stability of the candidates, we rely on the DFT phase stability by taking into account all phases that have been computed in a high-throughput fashion in the OQMD~\cite{saal_materials_2013, kirklin_oqmd_2015}. The database includes both experimentally observed phases reported in the ICSD as well as a wealth of hypothetical compounds, created by new decorations of frequently observed structural prototypes from the ICSD. The energies of all phases are subsequently used to construct the convex hulls for each of the X--Y--Z systems. The Gibbs triangle convex hull for the Ba--As--S system is shown in Fig.~\ref{fig:MLTernary}(c), and analogous figures for all other  ternary systems can be found in the Supplementary Materials.

In several systems we find that either \ce{X4Y2Z} or \ce{X5Y2Z2}, and in some cases even both compositions, are predicted to be thermodynamically stable. The only exceptions are Ba--Bi--S, Ba--Bi--Se,  Ba--Sb--S, Ca--Bi--S, Ca--Bi--Se, Ca--Sb--S, Sr--Bi--S, Sr--Bi--Se, Sr--Sb--S, and all magnesium containing compounds, Mg--Y--Z. Hence, phases where the Y and Z elements are two or more periods apart have a tendency to be unstable. 
Further, we notice a general trend that \ce{X4Y2Z} compounds are more stable than their \ce{X5Y2Z2} counterparts, i.e., a  \ce{X5Y2Z2} compound only lies on the convex hull of stability if also the corresponding \ce{X4Y2Z} compound is thermodynamically stable. The only exception is \ce{Ca5Sb2Se2} which is sufficiently low in energy to push  \ce{Ca4Sb2Se} away from the hull. Therefore, we henceforth focus on the characterization of the \ce{X4Y2Z} phases only.

Many of the oxide compounds \ce{X4Y2O} have been experimentally reported, and are known since the early 1970's.  \ce{Ca4Y2O}~\cite{hurng_alkaline-earth-metal_1989,hadenfeldt_darstellung_1988,xia_existence_2007}, \ce{Sr4Y2O}~\cite{hadenfeldt_darstellung_1991,wied_crystal_2014,klos_ternare_nodate}, and \ce{Ba4Y2O}~\cite{hadenfeldt_darstellung_1991,wied_crystal_2014,rohr_crystal_2010} with  Y=\{P, As, Sb, Bi\} crystallize all in a \ce{K2MgF4} structure type with $I4/mmm$ symmetry (some with minor distortions). Our MHM simulations recover this structure type for most of the oxides, or slight distortions thereof, giving us confidence that the structural searches are well converged. For the sulfides, selenides and tellurides, however, we find a diverse set of ground state structures for the distinct chemical compositions (the putative ground state structures at every composition can be found in the Supplemental Materials. This difference in behavior is expected due to the very strong metal-oxide bonding (giving rise to well defined, deep thermodynamic wells around the ground states) compared to the weaker metal-sulfide, metal-selenide, and metal-telluride interactions which lead to a higher density of configurational states in sulfides, selenides and tellurides.

\subsection{Structure and Bonding} 

One particular structure with \textit{I-42d} symmetry appears especially frequently in our search. In fact, the majority of the phases crystallize in this structure, which is shown in Fig.~\ref{fig:Structure}. Here, we use \ce{Ba4As2S} as a representative prototype to discuss the detailed structural features on behalf of all ground state phases in this particular structure type. The tetragonal structure has cell dimensions of  $a=b= 9.963$~\AA\, and $c= 10.021$~\AA. The Ba cation occupies the $16e$ Wyckoff position at $(0.43407, 0.81113, 0.43683)$, while the As and S atoms occupy the $8d$ and $4a$ sites at $(0.62758, \sfrac{1}{4}, \sfrac{1}{8})$ and $( 0, 0, 0)$, respectively. Each S atom has four nearest neighbors with a Ba--S  bond length of 3.231~\AA, which is slightly larger than in \ce{BaS} with the rocksalt structure (3.1935~\AA). On the other hand, each As atom has four Ba atoms in close proximity, two at a distance of 3.242~\AA,  and two at 3.265~\AA. These values compare well with the bond distance of 3.292~\AA\, in the experimentally reported \ce{Ba4As_{2.67}} compound with an anti-\ce{Th3P4} structure type. The four Ba atoms surrounding each cation form strongly distorted, almost flat tetrahedra, which are shown as yellow and blue polyhedra in the panels (a), (b) and (c) of Fig.~\ref{fig:Structure}. These tetrahedra are linked by sharing corners to form a network structure.

The \textit{I-42d} structure of \ce{Ba4As2S} is closely related to the experimentally reported, cubic \ce{Ba4As_{2.67}} phase with \textit{I-43d} symmetry~\cite{li_valence_2003}. Note that Li~\textit{et al.} classify \ce{Ba4As_{2.67}} as a Zintl compound~\cite{li_valence_2003}, with isolated As anions~\cite{janiak_riedel_2012}, but another way to interpret this compound is as a simple salt: A nominal composition assuming \ce{Ba^2+} \ce{As^3-} would be \ce{Ba3As2}, but the conventional cell contains 16~Ba atoms, resulting more conveniently in (\ce{Ba4As_{2.67}})$_4$. Li~\textit{et al.} interpret the structure as Ba atoms forming  two enantiomeric networks, isostructural to the ones shown in panel (d) of Fig.~\ref{fig:Structure}. The As atoms on the other hand occupy the total 12 combined As \textit{and} S sites of the corresponding \ce{Ba4As2S} structure in the voids of the Ba network. These sites however only have \sfrac{8}{9} occupancy (0.866 in experiment) to achieve the correct charge balance. In other words, the \ce{X4Y2Z} compounds can be interpreted as \ce{Th3X4} in the inverse \ce{Th3P4}-type structure, where the average charge of -2.67 of the anions is achieved by substituting the Th-site with two and one elements of mixed oxidation states -2 and -3 (here, Y and Z), respectively. Hence, the \ce{X4Y2Z} compounds are mixed-anion analogs of \ce{Ba4As_{2.67}}. These anion substitutions result in the tetragonal distortion away from the cubic symmetry.

\begin{figure}[!htbp]
\centering
\includegraphics[width=1\columnwidth]{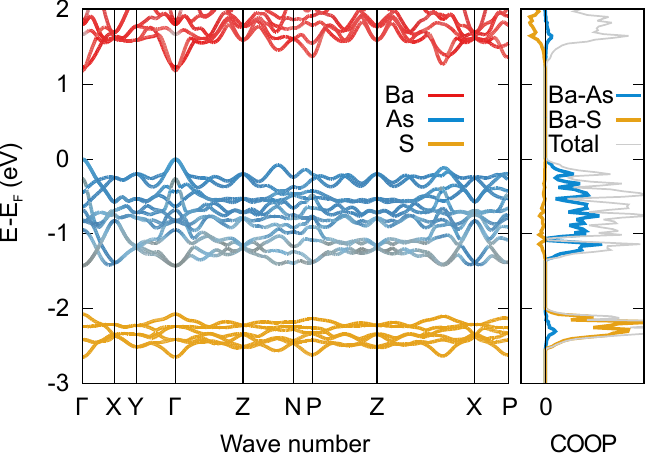}
\caption{
The electronic  structure of \ce{Ba4As2S}. The band structure on the left is colored based on the band character from atomic projections, while the Crystal Orbital Overlap Population (COOP)  is shown on the right. The grey line denoted with ``Total'' represents the total density of states (DOS).}\label{fig:BandsDOS}
\end{figure}

Following the classification of  Li~\textit{et al.}, we can also interpret the \ce{X4Y2Z} compounds as Zintl (valence) phases with isolated anions Y and Z  (according to the stoichiometry and assumed oxidation states),  where the alkaline earth metals donate their valence electrons to the semimetals (here the pnictogens and chalcogens) to satisfy the octet rule. Hence, we can expect semiconducting/insulating properties throughout. Fig.~\ref{fig:BandsDOS} shows the electronic band structure of \ce{Ba4As2S} as a representative prototype, with colors indicating the band character based on atomic projections. The band gap is direct, with the valence band maximum (VBM) and conduction band minimum (CBM) located at $\Gamma$. The valence bands stem dominantly from As $p$-type orbitals, with slight contributions from Ba with $p$ and $d$-type character. The S $p$ states are buried further below the Fermi level, at around -2.5~eV. The conduction bands stem primarily from the Ba $d$ states.

\begin{figure}[!htbp]
\centering
\includegraphics[width=1\columnwidth]{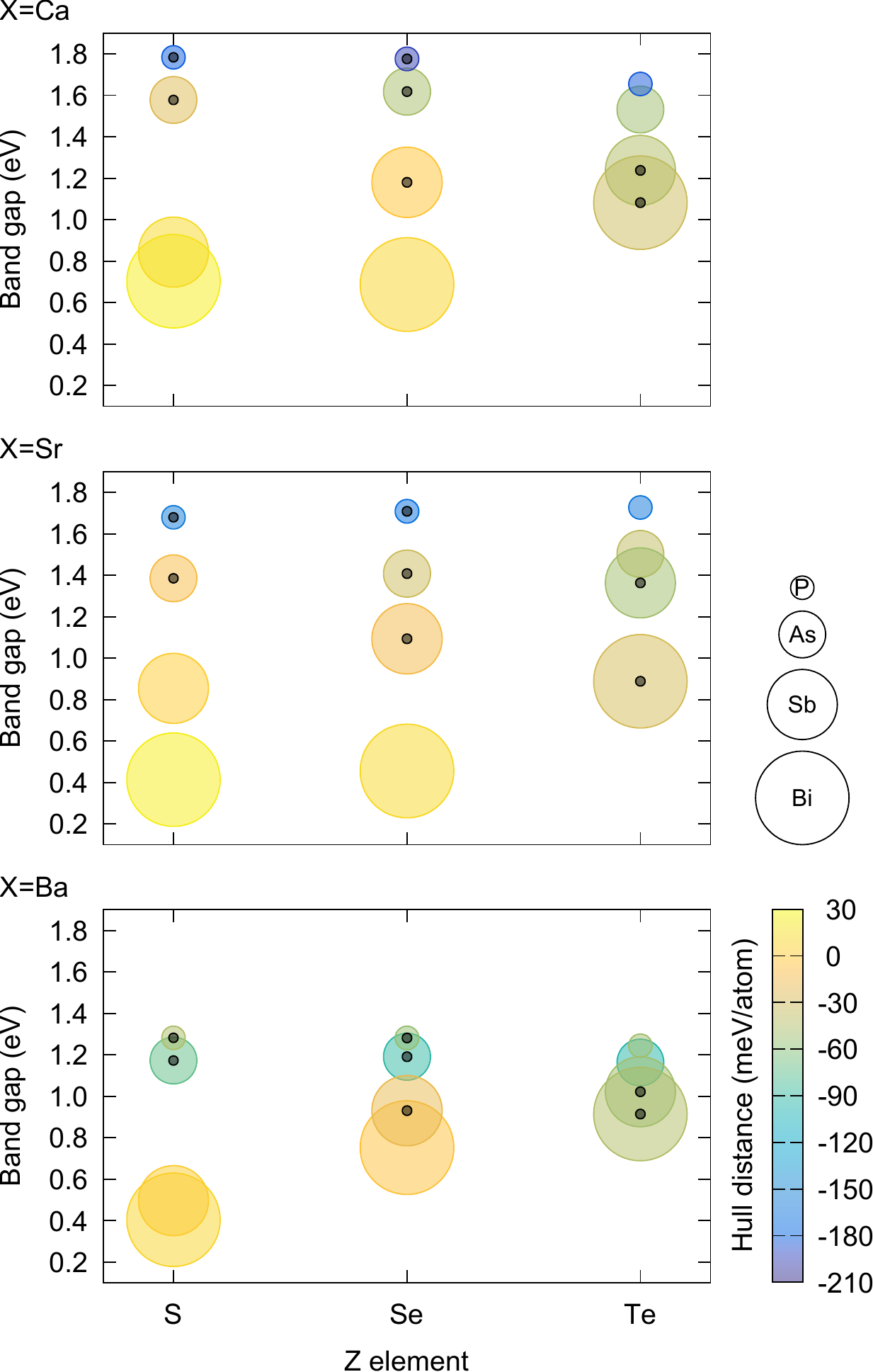}
\caption{The band gaps of the \ce{X4Y2Z} compounds computed with the PBE functional. The three panels represent values for X=Ca, Sr and Ba. In every panel, the x-axis corresponds to the Z elements, here \{S, Se, Te\}. The y-values of the center of every circle corresponds to the band gap maginitude, while the diameter represents the chemical element on the Y-site, here \{P, As, Sb, Bi\}. The color of the circles indicates the stability of the ground state structures at every composition (i.e., the convex hull distance), taking into account all relevant phases available in the OQMD. Finally, the presence of a black, filled dot at the center of the circles indicates that the ground state crystallizes in the tetragonal \textit{I-42d} structure.}\label{fig:Gaps}
\end{figure}

We do not expect strong (covalent) bonding in these  \textit{I-42d} structures, but a predominantly ionic behavior. Nevertheless, we inspect the orbital interactions via the COOP, since they will still play a role to some extent. In the right panel of Fig.~\ref{fig:BandsDOS} we show the COOP for the \ce{Ba4As2S} compound computed with the ADF Band package~\cite{te_velde_precise_1991,wiesenekker_quadratic_1991,franchini_becke_2013,franchini_accurate_2014}. These COOP are obtained by directly analyzing the contributions from the  LCAO basis set, which are employed by the ADF Band package. In contrast to plane wave codes, no atomic projections are required in ADF Band, and therefore one can expect more accurate results. The bonding behavior is very similar to the one for \ce{Sr4Bi3} in the cubic anti-\ce{Th3P4} structure type reported by Li~\textit{et al.}~\cite{li_valence_2003}, however with a shifted Fermi energy: here, the bonding states are completely filled below the Fermi level, while the anti-bonding states lie above the band gap. Again, note how the valence bands stem predominantly from the Ba--As interactions, while the Ba--S leads only to significant bonding contributions at energies of $-2$~eV.

\begin{figure*}[!htbp]
\centering
\includegraphics[width=0.7\textwidth]{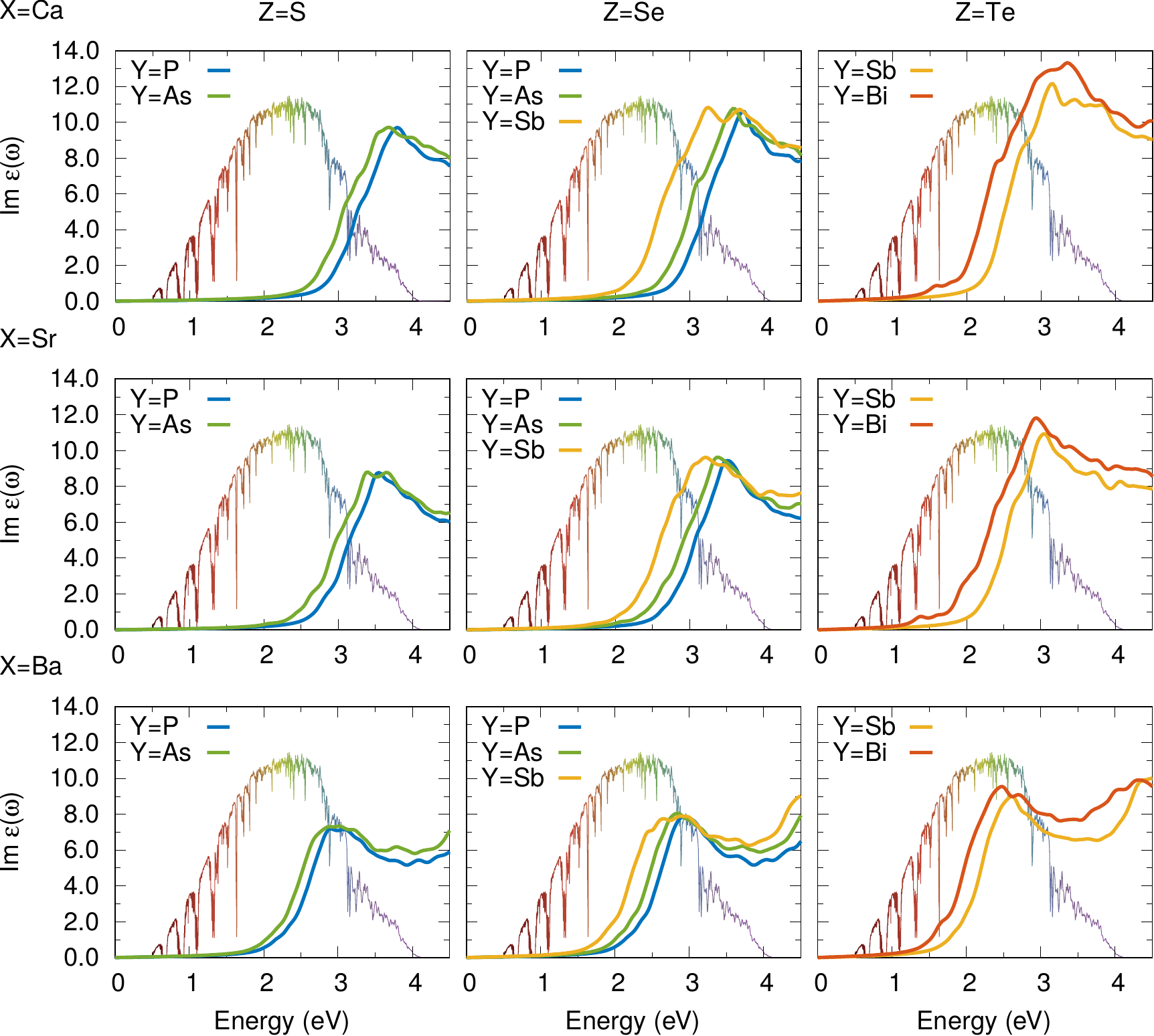}
\caption{The absorption spectra of the \ce{X4Y2Z} compounds in the \textit{I-42d} structure. The imaginary part of the dielectric function $\epsilon$ is shown with respect to the incident photon energy on top of the solar spectral irradiance~\cite{AM1.5}}\label{fig:Absorption}
\end{figure*}

Although the overall features of the band structure are very similar for all \ce{X4Y2Z} compounds (e.g., they all exhibit direct gaps at $\Gamma$), the magnitude of the band gaps changes significantly with the chemistry. Fig.~\ref{fig:Gaps} shows a graphical representation of the band gaps together with the thermodynamic stability. Each panel corresponds to an element for X, namely X=\{Ca, Sr, Ba\}. Within each panel, the y-axis defines the band gap, while the x-axis and the size of the circles indicates the element on the Z and Y sites, respectively. Finally, the color of the circles shows the hull distance, a measure of thermodynamic stability.

In the literature we find few reported synthesis of some of these ternary Zintl compounds. In an attempt to react \ce{Ca5Sb3}  with S, Hurng~\textit{et al.}~\cite{hurng_alkaline-earth-metal_1989} synthesized \ce{Ca4Sb_{2.4}S_{0.4}} in a defective cubic anti-\ce{Th3P4} structure type, very close to the nominal stoichiometry of \ce{Ca4Sb_{2}S}. However, based on our calculations, the ground state structure at the composition \ce{Ca4Sb_{2}S} is not the corresponding, ternary tetragonal \textit{I-42d} structure, but a monoclinic structure with \textit{P21/m} symmetry, which is energetically favorable by 27~meV/atom. Further, we see from the color scheme in Fig.~\ref{fig:Gaps} that  \ce{Ca4Sb_{2}S}, even in its monoclinic structure, is thermodynamically unstable with respect to decomposition into competing phases, as indicated by the positive hull distance of 12~meV/atom. Hulliger~\cite{hulliger_new_1979} reported the synthesis of ternary europium compounds in the inverse \ce{Th3P4} structure type (\ce{Eu4Y2Z} with Y=\{P, As, Sb, Bi\} and X=\{S, Se, Te\}) as well as \ce{Ca4Bi2Te}, \ce{Sm4Bi2Te} and \ce{Yb4Bi2Te}.

Based on these findings we can draw two conclusions: First, although the DFT formation energies at 0~K allows us to assess if a certain phase is thermodynamically stable or not, it does not uniquely provide us with an answer if it is synthesizeable, especially at finite temperatures. In particular,  \ce{Ca4Sb2S} is thermodynamically unstable at 0~K based on DFT, but nevertheless experimentally synthesizeable. Second, some phases that we determine to be unstable in the tetragonal \textit{I-42d} structure might in fact be accessible experimentally, stabilized due to entropic effects and structural disorder. E.g., the energetic penalty of $\Delta E=39$~meV/atom between the \textit{I-42d} phase of \ce{Ca4Sb2S} and the convex hull corresponds to a temperature of $T=\Delta E/k_B=452$~K, an energy difference that can be readily overcome at synthesis conditions.

We also identify three general chemical trends. First, the band gap decreases as we move down in the periodic table for the X site. Similarly, the  band gap decreases as we move down in the periodic table for the Y site. This behavior can be directly attributed to the decreasing electronegativity of the Y elements: the valence bands, which stem from their corresponding $p$ states, are pushed up in energy with respect to the conduction bands.  Finally, there is a less prominent tendency of converging band gap values as we move down in the periodic table for the Z site. Most importantly, we see that understanding these trends from chemical substitutions on the three distinct sites allows the tuning of the gap in a range between 0.3 and 1.8~eV.

Note that we analyze here the band gaps using the semi-local PBE functional, which is known to systematically underestimate the true gap energies. However, we also neglect the relativistic effects of spin-orbit coupling (SOC), which overall lowers the gap energies, especially for the compounds containing very heavy elements like bismuth (see Supplemental Materials). Hence, the two effects somewhat cancel each other out. Further, we show PBE results for consistency, since the training data for the ML models are trained on   data generated within the same DFT footing.

\subsection{Properties for Energy Applications}

As semiconductors, this range of band gaps is especially of interest for energy applications, and we will specifically discuss two of them in detail, namely photovoltaics and thermoelectrics. For the former, the Shockley-Queisser limit determines the ideal gap for a single  p-n junction photovoltaic cell at a value of 1.34~eV, resulting in a maximal (theoretical) efficiency of 33.7\%. Further, the direct nature of the band gap (both the VBM and the CBM are located at $\Gamma$, see Fig.~\ref{fig:BandsDOS}) would allow for an efficient photo absorption, one of the major drawbacks of silicon solar cells that currently dominate the photovoltaic market (diamond silicon has an indirect band gap and hence requires phonon-assisted photo-absorption).

\begin{figure*}[!htbp]
\centering
\includegraphics[width=0.7\textwidth]{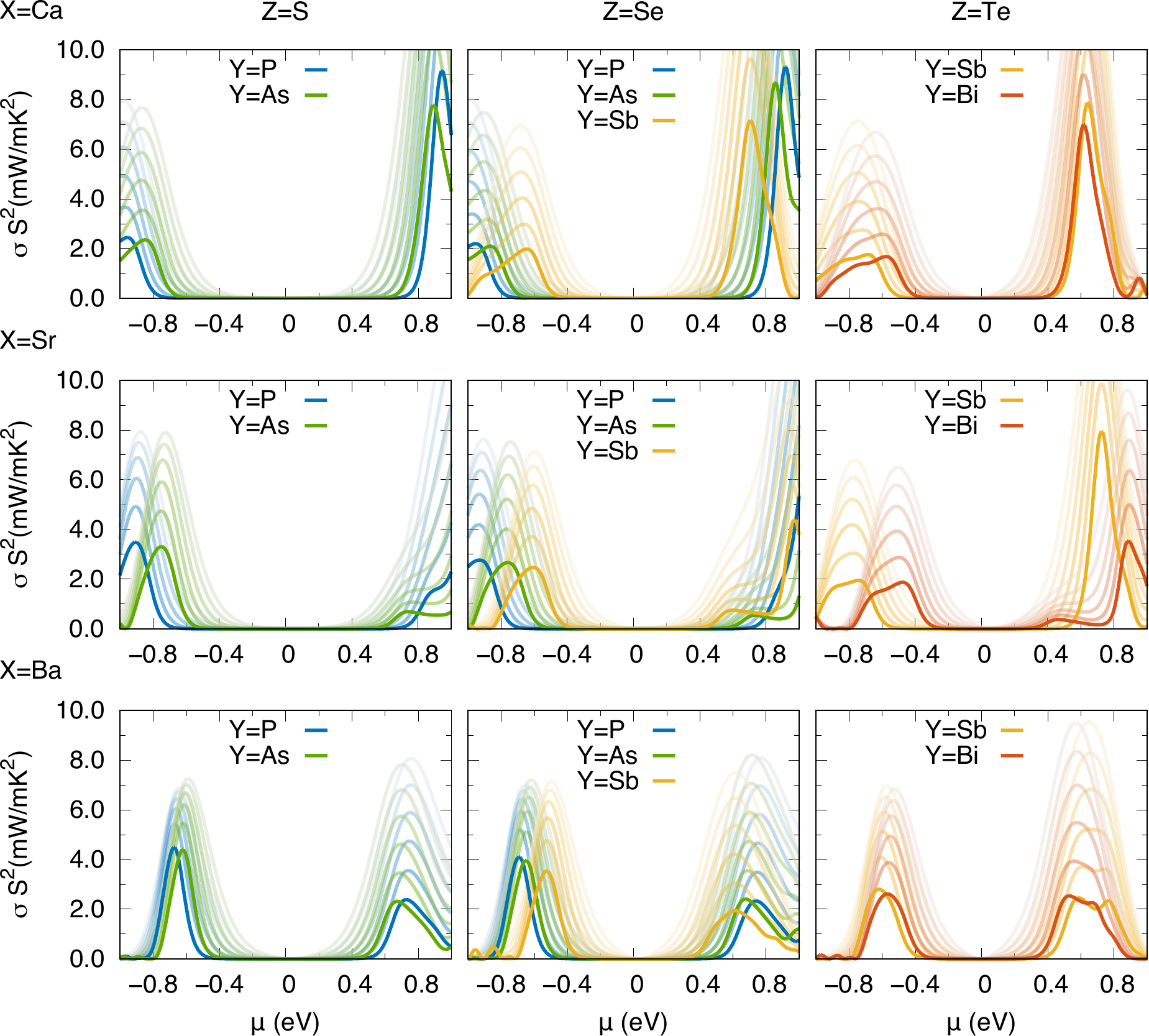}
\caption{The power factors $\sigma S^2$ of the \ce{X4Y2Z} compounds in the \textit{I-42d} structure as a function of temperature and electronic chemical potential $\mu$. Solid, dark lines represent values computed at a temperature of 300~K,  while the color gradient shows the results at higher temperatures in intervals of 100~K. The maximal temperature (lightest color) corresponds to 800~K.}\label{fig:PF}
\end{figure*}

To assess how well the \ce{X4Y2Z} compounds are suited for photovoltaic applications, we compute their absorption spectra and compare them with the solar spectrum. Semi-local functionals frequently underestimate the band gap, thereby also underestimating the absorption edge. We therefore compute the absorption using the hybrid HSE06 functional, which has been shown to improve the band gaps for many materials classes~\cite{garza_predicting_2016}. In Fig.~\ref{fig:Absorption} we show the imaginary part of the frequency dependent dielectric function, which is computed on a $6\times6\times6$ $k$-point mesh. Although this mesh might seem rather coarse, we performed careful convergence tests and determined that the overall features of the spectrum are well captured. In particular, we compare our results to $G_0W_0$ results and by solving the Bethe-Salpeter equation (see Supplemental Materials). Note, however, that we did not include the effect of spin-orbit coupling, which lowers the band gaps, especially for the bismuth-containing compounds (see Supplemental Materials).

As expected, the compounds that exhibit the best absorption behavior over a wide photon energy range in the solar spectrum are the ones with overall low band gaps, mainly containing Te. The best candidates are \ce{X4Y2Te} and \ce{X4Sb2Se}, with an absorption edge in the range of 1.2-2.0~eV and a high absorption peak around 1.8-3.0~eV, well below the UV regime. Hence, these materials might be well suited for photovoltaic applications, especially as thin films due their direct band gaps and the associated high absorption efficiency.

\begin{figure*}[!htbp]
\centering
\includegraphics[width=0.7\textwidth]{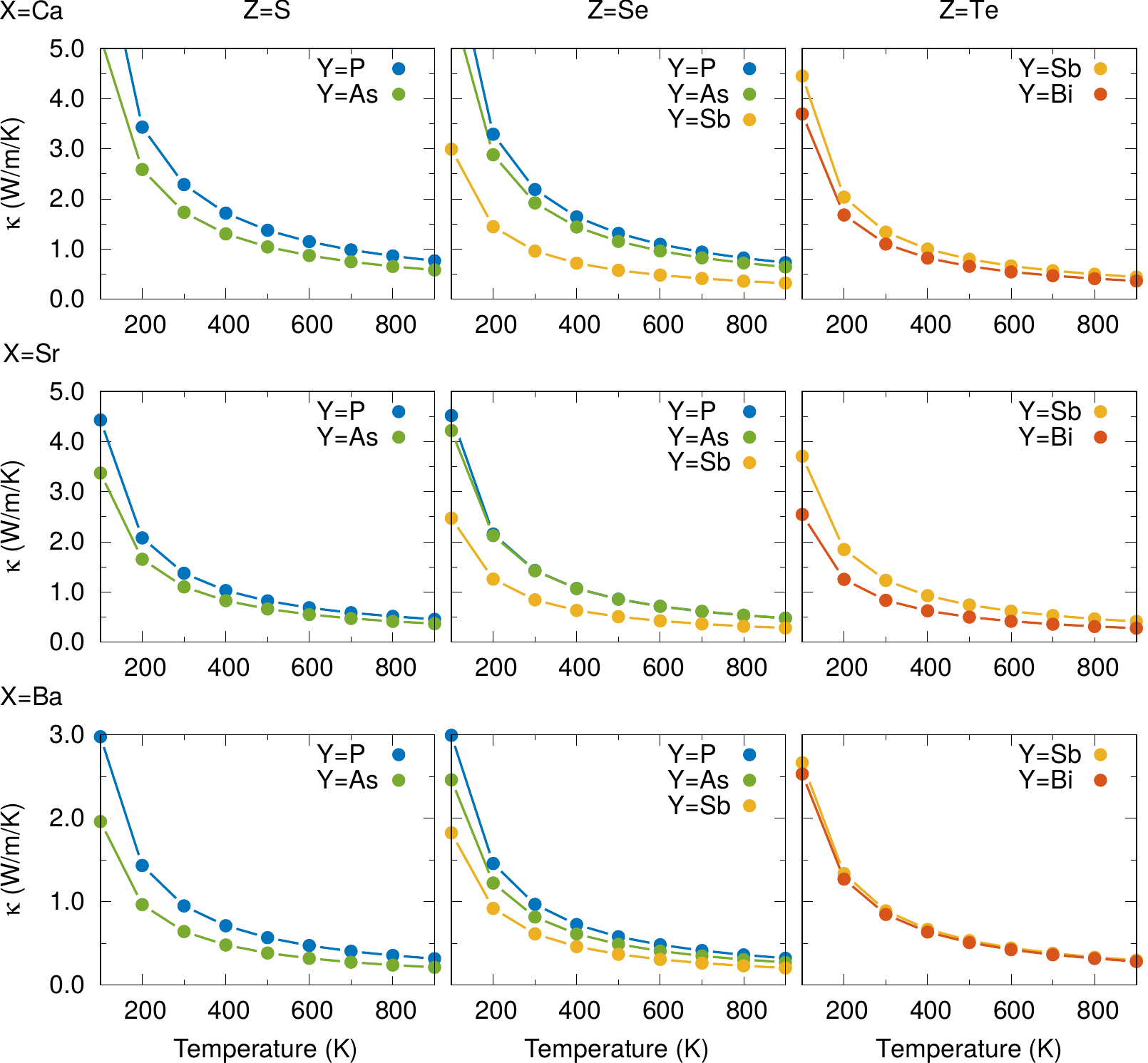}
\caption{The lattice thermal conductivities  $\kappa$ of the \ce{X4Y2Z} compounds in the \textit{I-42d} structure as a function of temperature. Note that at very high temperatures, the validity of the  Boltzmann transport equation breaks down as the phonon mean free paths become shorter than the smallest interatomic distances.}\label{fig:Kappa}
\end{figure*}

Zintl phases have been recently intensely studied as promising candidates for thermoelectric applications~\cite{kauzlarich_zintl_2007,snyder_complex_2008}. In general, the Carnot conversion efficiency of thermoelectric materials is  governed by the figure of merit $ZT=S^2 \sigma T/(\kappa_\text{e}+\kappa_{\text{L}})$, where $T$ is the absolute temperature, $S$ is the thermopower, $\sigma$ is the electrical conductivity, while $\text{e}$ and $\kappa_{\text{L}}$ are the  electronic and lattice thermal conductivities, respectively~\cite{rowe2005thermoelectrics}. Optimal values for the power factor $\sigma S$ (PF) can be achieved in heavily doped semiconductors with a carrier concentration in the range of $1\times 10^{19}$~cm$^{-1}$~\cite{zhao_high_2013,yin_optimization_2016}, with a narrow band gap which gives rise to large values of $\sigma$, while a sharp increase in the DOS around the Fermi level leads to an enhanced $S$ according to Mott's theory~\cite{rao_properties_2006}. Due to their chemical and structural complexity, Zintl compounds exhibit inherently low lattice thermal conductivities together with favorable power factors~\cite{toberer_phonon_2011,toberer_zintl_2010,zevalkink_thermoelectric_2014}. For example, the so-called  9-4-9 system represents Zintl compounds with \ce{A9M_{4+x}Pn9} stoichiometry (A=\{Yb, Eu, Ca, Sr\}; M=\{Mn, Zn, Cd\},   Pn=\{Sb, Bi\})  and was originally discovered in the 1970s, but only recently has it drawn attention for thermoelectric applications~\cite{bobev_probing_2004}. Through careful tuning of the carrier concentration  in the inexpensive \ce{Ca9Zn_{4+x}Sb9} compound, a high value of $zT$ has been recently reported up to $zT=1.1$ at 875~K~\cite{ohno_achieving_2017}.

In terms of the electronic properties, the band structure of \ce{Ba4As2S} in Fig.~\ref{fig:BandsDOS} shows that many local extrema exist in a small energy range below the Fermi level at low-symmetry $k$-points. These degenerate energy levels give rise to the sharp increase in the DOS around the Fermi energy. $p$-doping could be readily achieved either through partially substituting the Y-site with group 13 or 14 elements, or by carefully tuning the Y/Z ratio on the anion sites towards values larger than 2. Similar arguments apply for the other  \ce{X4Y2Z} compounds, where band degeneracies are found not only for the valence but also the conduction bands (e.g., \ce{Ca4Sb2Te},  see Supplemental Materials). Analogously,  $n$-doping can be achieved by tuning the Y/Z ratio towards values below 2. The experimentally reported room-temperature Seebeck coefficients of 140 and 230~$\mu$V/K (both p-type) for \ce{Eu4As2Te}  and \ce{Sm4Bi2Te} in the anti-\ce{Th3P4} structure type~\cite{hulliger_new_1979} indicate that similar values can be achieved for the compounds studied in the present work.

Within the relaxation-time approximation of the electronic Boltzmann equation, the electrical conductivity $\sigma$ and the electronic thermal conductivity $\kappa_\text{e}$ depend in the constant relaxation time $\tau$. An accurate  computation of $\tau$ from first principles is challenging, and depends on the scattering mechanisms with phonons, carriers, defects, and grain boundaries. Nevertheless, we can gain some (qualitative) insight without explicitly computing  $\tau$, but by approximating  it  with values commonly observed in doped semiconductors, in the range of $10^{-14}$~s. Here, we use $\tau=3.4\times 10^{-14}$~s  throughout, a value which was initially obtained by Bilc~\textit{et al.} by fitting the electrical resistivity to experimental data for \ce{Fe2VAl_{1-x}M_{x}},  M =\{Si, Ge\}~\cite{bilc_low-dimensional_2015}. The electronic bands are resolved on a dense $35\times35\times35$ $k$-point mesh, with the PBE exchange-correlation functional. Fig.~\ref{fig:PF} shows the power factors $\sigma S^2$ computed with the \texttt{Boltztrap} package as a function of doping and temperature. The values of $\sigma S^2$ range between 2 and 5~mWm$^{-1}$K$^{-2}$ at 300~K and $p$-doping, comparable to other state-of-the art thermoelectric materials, while values of up to 10~mWm$^{-1}$K$^{-2}$ are predicted for $n$-doping. These values further increase with temperatures that are close to operating conditions of thermoelectric generators.

Due to the above-mentioned issues in accurately estimating $\tau$, it is difficult to directly determine the figure of merit $ZT$. However, we can use the Wiedemann-Franz law to rewrite $ZT$ in terms of the Lorentz factor $L=\kappa_e/T \sigma$ by factorizing $ZT$ into a purely electronic part, $S^2/L$, and the electronic fraction of the total thermal conductivity, $\frac{\kappa_e}{\kappa_\text{e}+\kappa_\text{L}}$
\begin{align*}
ZT=\frac{S^2}{L}\frac{\kappa_e}{\kappa_\text{e}+\kappa_\text{L}} < \frac{S^2}{L}\eqqcolon ZT_e.
\end{align*}
In this way, the relaxation time cancels between the numerators and denominators, assuming that $\tau$ is wave vector independent. If we additionally neglect the effect of $\kappa_\text{L}$, the value of $ZT_e=S^2/L$ provides a convenient measure for the upper limit of $ZT$. Fig.~S1 in the Supplemental Materials  shows $ZT_e$ as a function of doping and temperature for all compounds with the \textit{I-42d} symmetry ground state structure. Overall, the results show that values of $ZT\approx 1$ can be achieved with a careful doping in the appropriate regime.

In terms of lattice dynamics, Fig.~S2 in the Supplemental Materials shows the phonon dispersion and the partial phonon DOS of \ce{Ba4As2S} as a representative prototype, where every band is colored according to the amplitudes of the atomic eigendisplacements. There are three clearly distinct frequency regimes that are separated by pseudo gaps in the DOS. (a) The very low frequencies up to about 2~THz are dominated by Ba vibrations (on the X-sites), with some mixing of  As (Y-site) contributions at 2-3~THz. (b) Between 4 and 5~THz we observe phonon bands that stem mainly from As (Y-site) vibrations. (c) The highest frequency modes at above 5~THz stem almost entirely from the S (Z-site) vibrations.

In addition to the phonon spectrum, we require higher order anharmonic contributions to the potential energy which give rise to phonon-phonon interaction to assess the lattice thermal conductivity $\kappa_\text{L}$. 
Using the recently developed CSLD technique, we fit the third order force constants, and use them to solve the linearized Boltzmann phonon transport equation using the \texttt{ShengBTE} package. The resulting lattice thermal conductivities are shown in Fig.~\ref{fig:Kappa} as a function of temperature.

Typical for Zintl compounds, the values of  $\kappa_\text{L}$ are overall very low, and less than $\approx 2$~Wm$^{-1}$K$^{-1}$ for all compounds at 300~K. We notice two general chemical trends:
\begin{enumerate}
	\item  $\kappa_\text{L}$ decreases with the increasing mass of the X-element. This behavior can be readily attributed to the overall decrease in the vibrational frequencies of the acoustic modes, which carry the majority of the heat. The phonon dispersion and the partial phonon DOS in Fig.~S2 (Supplemental Materials) shows that the low-frequency modes are dominated by the vibration of the X-elements, hence it is not surprising that it strongly affects the thermal conductivity.
	\item Given fixed X and Z elements, the value of  $\kappa_\text{L}$ decreases with increasing mass of the Y-element. We can attribute this correlation to the strong influence of the Y-elements on the low-frequency, acoustic phonons: the phonon band structures shows a mixing of  Y and X-element vibrations. The higher the mass of the Y element, the stronger the mixing.
\end{enumerate}

The lowest thermal conductivity is observed for \ce{Ba4Sb2Se} with $\kappa_\text{L}=0.61$~Wm$^{-1}$K$^{-1}$ at 300~K, which is close to the value of the best currently known thermoelectric material, \ce{SnSe} ($\kappa_\text{L}=0.47$~Wm$^{-1}$K$^{-1}$)\cite{SnSe,doi:10.1038/nphys3492}. Note that we only report the intrinsic, bulk thermal conductivities here. Doping and structural engineering through nano structuring can further reduce the values of $\kappa$. In fact, we expect that, when synthesized, the anionic sites might exhibit some degree of disorder, which could further reduce the thermal conductivity while preserving the favorable electronic properties.

\section{Conclusions}

In summary, we present the properties of a class of ternary semiconducting compounds, which we discovered by effectively leveraging a machine learning model to predict the composition/band gap and a structure prediction scheme to assess their underlying ground states. Using \textit{ab initio} calculations, we study in detail the properties of this  class of materials with \ce{X4Y2Z} composition. The most prominent ground state structure that we identify has \textit{I-42d} symmetry and is closely related to the binary anti-\ce{Th3P4} structure type with cubic symmetry. 

Our calculations show that the electronic structure of these compounds can be tuned through chemical substitution, leading to band gaps that are suited for a wide range of energy applications. The direct band gaps and the favorable absorption spectra of \ce{X4Y2Te} and \ce{X4Sb2Se} are especially promising for photovoltaic applications. Further, the high band degeneracy in the valence/conduction bands lead to high power factors, which, together with the low lattice thermal conductivities, renders these materials potentially attractive in thermoelectric generators. 

\section{Acknowledgments}\label{sec:ack}
We thank Prof. R. Hoffmann for valuable expert discussions. M.A. (DFT calculations) acknowledges support from the Novartis Universit{\"a}t Basel Excellence Scholarship for Life Sciences and the Swiss
National Science Foundation (project No. P300P2-158407, P300P2-174475). M.G.G. (COOP calculations) acknowledges support from the Rubicon Research Programme (project 019.161BT.031), which is (partly) financed by the Netherlands Organization for Scientific Research (NWO). 
L.W. (ML model) and C.W. acknowledge the financial assistance Award 70NANB14H012 from the US Department of Commerce, National Institute of Standards and Technology as part of the Center for Hierarchical Materials Design (CHiMaD).
V.I.H  (OQMD calculations) acknowledges support from the National Science Foundation Materials Research Science
and Engineering Center (MRSEC) of Northwestern University (DMR-1720139).
The computational resources from the Swiss National Supercomputing Center in Lugano (projects s621,s700, and s861),
the Extreme Science and Engineering Discovery Environment (XSEDE) (which is supported by National Science
Foundation grant number OCI-1053575), the Bridges system at the Pittsburgh Supercomputing Center (PSC) (which is
supported by NSF award number ACI-1445606), the Quest high performance computing facility at Northwestern University,
and the National Energy Research Scientific Computing Center (DOE: DE-AC02-05CH11231), are gratefully acknowledged.



\begin{thebibliography}{87}%
\makeatletter
\providecommand \@ifxundefined [1]{%
 \@ifx{#1\undefined}
}%
\providecommand \@ifnum [1]{%
 \ifnum #1\expandafter \@firstoftwo
 \else \expandafter \@secondoftwo
 \fi
}%
\providecommand \@ifx [1]{%
 \ifx #1\expandafter \@firstoftwo
 \else \expandafter \@secondoftwo
 \fi
}%
\providecommand \natexlab [1]{#1}%
\providecommand \enquote  [1]{``#1''}%
\providecommand \bibnamefont  [1]{#1}%
\providecommand \bibfnamefont [1]{#1}%
\providecommand \citenamefont [1]{#1}%
\providecommand \href@noop [0]{\@secondoftwo}%
\providecommand \href [0]{\begingroup \@sanitize@url \@href}%
\providecommand \@href[1]{\@@startlink{#1}\@@href}%
\providecommand \@@href[1]{\endgroup#1\@@endlink}%
\providecommand \@sanitize@url [0]{\catcode `\\12\catcode `\$12\catcode
  `\&12\catcode `\#12\catcode `\^12\catcode `\_12\catcode `\%12\relax}%
\providecommand \@@startlink[1]{}%
\providecommand \@@endlink[0]{}%
\providecommand \url  [0]{\begingroup\@sanitize@url \@url }%
\providecommand \@url [1]{\endgroup\@href {#1}{\urlprefix }}%
\providecommand \urlprefix  [0]{URL }%
\providecommand \Eprint [0]{\href }%
\providecommand \doibase [0]{http://dx.doi.org/}%
\providecommand \selectlanguage [0]{\@gobble}%
\providecommand \bibinfo  [0]{\@secondoftwo}%
\providecommand \bibfield  [0]{\@secondoftwo}%
\providecommand \translation [1]{[#1]}%
\providecommand \BibitemOpen [0]{}%
\providecommand \bibitemStop [0]{}%
\providecommand \bibitemNoStop [0]{.\EOS\space}%
\providecommand \EOS [0]{\spacefactor3000\relax}%
\providecommand \BibitemShut  [1]{\csname bibitem#1\endcsname}%
\let\auto@bib@innerbib\@empty
\bibitem [{\citenamefont {Saal}\ \emph {et~al.}(2013)\citenamefont {Saal},
  \citenamefont {Kirklin}, \citenamefont {Aykol}, \citenamefont {Meredig},\
  and\ \citenamefont {Wolverton}}]{saal_materials_2013}%
  \BibitemOpen
  \bibfield  {author} {\bibinfo {author} {\bibfnamefont {J.~E.}\ \bibnamefont
  {Saal}}, \bibinfo {author} {\bibfnamefont {S.}~\bibnamefont {Kirklin}},
  \bibinfo {author} {\bibfnamefont {M.}~\bibnamefont {Aykol}}, \bibinfo
  {author} {\bibfnamefont {B.}~\bibnamefont {Meredig}}, \ and\ \bibinfo
  {author} {\bibfnamefont {C.}~\bibnamefont {Wolverton}},\ }\href {\doibase
  10.1007/s11837-013-0755-4} {\bibfield  {journal} {\bibinfo  {journal}
  {{JOM}}\ }\textbf {\bibinfo {volume} {65}},\ \bibinfo {pages} {1501}
  (\bibinfo {year} {2013})}\BibitemShut {NoStop}%
\bibitem [{\citenamefont {Kirklin}\ \emph {et~al.}(2015)\citenamefont
  {Kirklin}, \citenamefont {Saal}, \citenamefont {Meredig}, \citenamefont
  {Thompson}, \citenamefont {Doak}, \citenamefont {Aykol}, \citenamefont
  {R{\"u}hl},\ and\ \citenamefont {Wolverton}}]{kirklin_oqmd_2015}%
  \BibitemOpen
  \bibfield  {author} {\bibinfo {author} {\bibfnamefont {S.}~\bibnamefont
  {Kirklin}}, \bibinfo {author} {\bibfnamefont {J.~E.}\ \bibnamefont {Saal}},
  \bibinfo {author} {\bibfnamefont {B.}~\bibnamefont {Meredig}}, \bibinfo
  {author} {\bibfnamefont {A.}~\bibnamefont {Thompson}}, \bibinfo {author}
  {\bibfnamefont {J.~W.}\ \bibnamefont {Doak}}, \bibinfo {author}
  {\bibfnamefont {M.}~\bibnamefont {Aykol}}, \bibinfo {author} {\bibfnamefont
  {S.}~\bibnamefont {R{\"u}hl}}, \ and\ \bibinfo {author} {\bibfnamefont
  {C.}~\bibnamefont {Wolverton}},\ }\href@noop {} {\bibfield  {journal}
  {\bibinfo  {journal} {npj Computational Materials}\ }\textbf {\bibinfo
  {volume} {1}},\ \bibinfo {pages} {15010} (\bibinfo {year}
  {2015})}\BibitemShut {NoStop}%
\bibitem [{\citenamefont {Jain}\ \emph {et~al.}(2013)\citenamefont {Jain},
  \citenamefont {Ong}, \citenamefont {Hautier}, \citenamefont {Chen},
  \citenamefont {Richards}, \citenamefont {Dacek}, \citenamefont {Cholia},
  \citenamefont {Gunter}, \citenamefont {Skinner}, \citenamefont {Ceder},\ and\
  \citenamefont {Persson}}]{Jain2013}%
  \BibitemOpen
  \bibfield  {author} {\bibinfo {author} {\bibfnamefont {A.}~\bibnamefont
  {Jain}}, \bibinfo {author} {\bibfnamefont {S.~P.}\ \bibnamefont {Ong}},
  \bibinfo {author} {\bibfnamefont {G.}~\bibnamefont {Hautier}}, \bibinfo
  {author} {\bibfnamefont {W.}~\bibnamefont {Chen}}, \bibinfo {author}
  {\bibfnamefont {W.~D.}\ \bibnamefont {Richards}}, \bibinfo {author}
  {\bibfnamefont {S.}~\bibnamefont {Dacek}}, \bibinfo {author} {\bibfnamefont
  {S.}~\bibnamefont {Cholia}}, \bibinfo {author} {\bibfnamefont
  {D.}~\bibnamefont {Gunter}}, \bibinfo {author} {\bibfnamefont
  {D.}~\bibnamefont {Skinner}}, \bibinfo {author} {\bibfnamefont
  {G.}~\bibnamefont {Ceder}}, \ and\ \bibinfo {author} {\bibfnamefont
  {K.}~\bibnamefont {Persson}},\ }\href {\doibase 10.1063/1.4812323} {\bibfield
   {journal} {\bibinfo  {journal} {APL Mater.}\ }\textbf {\bibinfo {volume}
  {1}},\ \bibinfo {pages} {011002} (\bibinfo {year} {2013})}\BibitemShut
  {NoStop}%
\bibitem [{\citenamefont {Curtarolo}\ \emph {et~al.}(2012)\citenamefont
  {Curtarolo}, \citenamefont {Setyawan}, \citenamefont {Wang}, \citenamefont
  {Xue}, \citenamefont {Yang}, \citenamefont {Taylor}, \citenamefont {Nelson},
  \citenamefont {Hart}, \citenamefont {Sanvito}, \citenamefont
  {Buongiorno-Nardelli}, \citenamefont {Mingo},\ and\ \citenamefont
  {Levy}}]{curtarolo_aflowlib.org_2012}%
  \BibitemOpen
  \bibfield  {author} {\bibinfo {author} {\bibfnamefont {S.}~\bibnamefont
  {Curtarolo}}, \bibinfo {author} {\bibfnamefont {W.}~\bibnamefont {Setyawan}},
  \bibinfo {author} {\bibfnamefont {S.}~\bibnamefont {Wang}}, \bibinfo {author}
  {\bibfnamefont {J.}~\bibnamefont {Xue}}, \bibinfo {author} {\bibfnamefont
  {K.}~\bibnamefont {Yang}}, \bibinfo {author} {\bibfnamefont {R.~H.}\
  \bibnamefont {Taylor}}, \bibinfo {author} {\bibfnamefont {L.~J.}\
  \bibnamefont {Nelson}}, \bibinfo {author} {\bibfnamefont {G.~L.~W.}\
  \bibnamefont {Hart}}, \bibinfo {author} {\bibfnamefont {S.}~\bibnamefont
  {Sanvito}}, \bibinfo {author} {\bibfnamefont {M.}~\bibnamefont
  {Buongiorno-Nardelli}}, \bibinfo {author} {\bibfnamefont {N.}~\bibnamefont
  {Mingo}}, \ and\ \bibinfo {author} {\bibfnamefont {O.}~\bibnamefont {Levy}},\
  }\href {\doibase 10.1016/j.commatsci.2012.02.002} {\bibfield  {journal}
  {\bibinfo  {journal} {Comput. Mater. Sci.}\ }\textbf {\bibinfo {volume}
  {58}},\ \bibinfo {pages} {227} (\bibinfo {year} {2012})}\BibitemShut
  {NoStop}%
\bibitem [{\citenamefont {Curtarolo}\ \emph {et~al.}(2013)\citenamefont
  {Curtarolo}, \citenamefont {Hart}, \citenamefont {Nardelli}, \citenamefont
  {Mingo}, \citenamefont {Sanvito},\ and\ \citenamefont
  {Levy}}]{curtarolo_high-throughput_2013}%
  \BibitemOpen
  \bibfield  {author} {\bibinfo {author} {\bibfnamefont {S.}~\bibnamefont
  {Curtarolo}}, \bibinfo {author} {\bibfnamefont {G.~L.~W.}\ \bibnamefont
  {Hart}}, \bibinfo {author} {\bibfnamefont {M.~B.}\ \bibnamefont {Nardelli}},
  \bibinfo {author} {\bibfnamefont {N.}~\bibnamefont {Mingo}}, \bibinfo
  {author} {\bibfnamefont {S.}~\bibnamefont {Sanvito}}, \ and\ \bibinfo
  {author} {\bibfnamefont {O.}~\bibnamefont {Levy}},\ }\href {\doibase
  10.1038/nmat3568} {\bibfield  {journal} {\bibinfo  {journal} {Nat. Mater.}\
  }\textbf {\bibinfo {volume} {12}},\ \bibinfo {pages} {191} (\bibinfo {year}
  {2013})}\BibitemShut {NoStop}%
\bibitem [{\citenamefont {Toher}\ \emph {et~al.}(2017)\citenamefont {Toher},
  \citenamefont {Oses}, \citenamefont {Hicks}, \citenamefont {Gossett},
  \citenamefont {Rose}, \citenamefont {Nath}, \citenamefont {Usanmaz},
  \citenamefont {Ford}, \citenamefont {Perim}, \citenamefont {Calderon},
  \citenamefont {Plata}, \citenamefont {Lederer}, \citenamefont {Jahnátek},
  \citenamefont {Setyawan}, \citenamefont {Wang}, \citenamefont {Xue},
  \citenamefont {Rasch}, \citenamefont {Chepulskii}, \citenamefont {Taylor},
  \citenamefont {Gomez}, \citenamefont {Shi}, \citenamefont {Supka},
  \citenamefont {Orabi}, \citenamefont {Gopal}, \citenamefont {Cerasoli},
  \citenamefont {Liyanage}, \citenamefont {Wang}, \citenamefont {Siloi},
  \citenamefont {Agapito}, \citenamefont {Nyshadham}, \citenamefont {Hart},
  \citenamefont {Carrete}, \citenamefont {Legrain}, \citenamefont {Mingo},
  \citenamefont {Zurek}, \citenamefont {Isayev}, \citenamefont {Tropsha},
  \citenamefont {Sanvito}, \citenamefont {Hanson}, \citenamefont {Takeuchi},
  \citenamefont {Mehl}, \citenamefont {Kolmogorov}, \citenamefont {Yang},
  \citenamefont {D'Amico}, \citenamefont {Calzolari}, \citenamefont {Costa},
  \citenamefont {De~Gennaro}, \citenamefont {Nardelli}, \citenamefont
  {Fornari}, \citenamefont {Levy},\ and\ \citenamefont
  {Curtarolo}}]{toher_aflow_2017}%
  \BibitemOpen
  \bibfield  {author} {\bibinfo {author} {\bibfnamefont {C.}~\bibnamefont
  {Toher}}, \bibinfo {author} {\bibfnamefont {C.}~\bibnamefont {Oses}},
  \bibinfo {author} {\bibfnamefont {D.}~\bibnamefont {Hicks}}, \bibinfo
  {author} {\bibfnamefont {E.}~\bibnamefont {Gossett}}, \bibinfo {author}
  {\bibfnamefont {F.}~\bibnamefont {Rose}}, \bibinfo {author} {\bibfnamefont
  {P.}~\bibnamefont {Nath}}, \bibinfo {author} {\bibfnamefont {D.}~\bibnamefont
  {Usanmaz}}, \bibinfo {author} {\bibfnamefont {D.~C.}\ \bibnamefont {Ford}},
  \bibinfo {author} {\bibfnamefont {E.}~\bibnamefont {Perim}}, \bibinfo
  {author} {\bibfnamefont {C.~E.}\ \bibnamefont {Calderon}}, \bibinfo {author}
  {\bibfnamefont {J.~J.}\ \bibnamefont {Plata}}, \bibinfo {author}
  {\bibfnamefont {Y.}~\bibnamefont {Lederer}}, \bibinfo {author} {\bibfnamefont
  {M.}~\bibnamefont {Jahnátek}}, \bibinfo {author} {\bibfnamefont
  {W.}~\bibnamefont {Setyawan}}, \bibinfo {author} {\bibfnamefont
  {S.}~\bibnamefont {Wang}}, \bibinfo {author} {\bibfnamefont {J.}~\bibnamefont
  {Xue}}, \bibinfo {author} {\bibfnamefont {K.}~\bibnamefont {Rasch}}, \bibinfo
  {author} {\bibfnamefont {R.~V.}\ \bibnamefont {Chepulskii}}, \bibinfo
  {author} {\bibfnamefont {R.~H.}\ \bibnamefont {Taylor}}, \bibinfo {author}
  {\bibfnamefont {G.}~\bibnamefont {Gomez}}, \bibinfo {author} {\bibfnamefont
  {H.}~\bibnamefont {Shi}}, \bibinfo {author} {\bibfnamefont {A.~R.}\
  \bibnamefont {Supka}}, \bibinfo {author} {\bibfnamefont {R.~A. R.~A.}\
  \bibnamefont {Orabi}}, \bibinfo {author} {\bibfnamefont {P.}~\bibnamefont
  {Gopal}}, \bibinfo {author} {\bibfnamefont {F.~T.}\ \bibnamefont {Cerasoli}},
  \bibinfo {author} {\bibfnamefont {L.}~\bibnamefont {Liyanage}}, \bibinfo
  {author} {\bibfnamefont {H.}~\bibnamefont {Wang}}, \bibinfo {author}
  {\bibfnamefont {I.}~\bibnamefont {Siloi}}, \bibinfo {author} {\bibfnamefont
  {L.~A.}\ \bibnamefont {Agapito}}, \bibinfo {author} {\bibfnamefont
  {C.}~\bibnamefont {Nyshadham}}, \bibinfo {author} {\bibfnamefont {G.~L.~W.}\
  \bibnamefont {Hart}}, \bibinfo {author} {\bibfnamefont {J.}~\bibnamefont
  {Carrete}}, \bibinfo {author} {\bibfnamefont {F.}~\bibnamefont {Legrain}},
  \bibinfo {author} {\bibfnamefont {N.}~\bibnamefont {Mingo}}, \bibinfo
  {author} {\bibfnamefont {E.}~\bibnamefont {Zurek}}, \bibinfo {author}
  {\bibfnamefont {O.}~\bibnamefont {Isayev}}, \bibinfo {author} {\bibfnamefont
  {A.}~\bibnamefont {Tropsha}}, \bibinfo {author} {\bibfnamefont
  {S.}~\bibnamefont {Sanvito}}, \bibinfo {author} {\bibfnamefont {R.~M.}\
  \bibnamefont {Hanson}}, \bibinfo {author} {\bibfnamefont {I.}~\bibnamefont
  {Takeuchi}}, \bibinfo {author} {\bibfnamefont {M.~J.}\ \bibnamefont {Mehl}},
  \bibinfo {author} {\bibfnamefont {A.~N.}\ \bibnamefont {Kolmogorov}},
  \bibinfo {author} {\bibfnamefont {K.}~\bibnamefont {Yang}}, \bibinfo {author}
  {\bibfnamefont {P.}~\bibnamefont {D'Amico}}, \bibinfo {author} {\bibfnamefont
  {A.}~\bibnamefont {Calzolari}}, \bibinfo {author} {\bibfnamefont
  {M.}~\bibnamefont {Costa}}, \bibinfo {author} {\bibfnamefont
  {R.}~\bibnamefont {De~Gennaro}}, \bibinfo {author} {\bibfnamefont {M.~B.}\
  \bibnamefont {Nardelli}}, \bibinfo {author} {\bibfnamefont {M.}~\bibnamefont
  {Fornari}}, \bibinfo {author} {\bibfnamefont {O.}~\bibnamefont {Levy}}, \
  and\ \bibinfo {author} {\bibfnamefont {S.}~\bibnamefont {Curtarolo}},\ }\href
  {http://arxiv.org/abs/1712.00422} {\bibfield  {journal} {\bibinfo  {journal}
  {arXiv:1712.00422 [cond-mat]}\ } (\bibinfo {year} {2017})},\ \bibinfo {note}
  {arXiv: 1712.00422}\BibitemShut {NoStop}%
\bibitem [{\citenamefont {Oganov}(2010)}]{oganov_modern_2010}%
  \BibitemOpen
  \bibfield  {author} {\bibinfo {author} {\bibfnamefont {A.~R.}\ \bibnamefont
  {Oganov}},\ }\href@noop {} {\emph {\bibinfo {title} {Modern {Methods} of
  {Crystal} {Structure} {Prediction}}}},\ \bibinfo {edition} {1st}\ ed.\
  (\bibinfo  {publisher} {Wiley-VCH Verlag GmbH \& Co. KGaA},\ \bibinfo {year}
  {2010})\BibitemShut {NoStop}%
\bibitem [{\citenamefont {Ward}\ and\ \citenamefont
  {Wolverton}(2017)}]{ward_atomistic_2017}%
  \BibitemOpen
  \bibfield  {author} {\bibinfo {author} {\bibfnamefont {L.}~\bibnamefont
  {Ward}}\ and\ \bibinfo {author} {\bibfnamefont {C.}~\bibnamefont
  {Wolverton}},\ }\href {\doibase 10.1016/j.cossms.2016.07.002} {\bibfield
  {journal} {\bibinfo  {journal} {Current Opinion in Solid State and Materials
  Science}\ }\bibinfo {series} {Materials {Informatics}: {Insights},
  {Infrastructure}, and {Methods}},\ \textbf {\bibinfo {volume} {21}},\
  \bibinfo {pages} {167} (\bibinfo {year} {2017})}\BibitemShut {NoStop}%
\bibitem [{\citenamefont {Seko}\ \emph {et~al.}(2014)\citenamefont {Seko},
  \citenamefont {Maekawa}, \citenamefont {Tsuda},\ and\ \citenamefont
  {Tanaka}}]{seko_machine_2014}%
  \BibitemOpen
  \bibfield  {author} {\bibinfo {author} {\bibfnamefont {A.}~\bibnamefont
  {Seko}}, \bibinfo {author} {\bibfnamefont {T.}~\bibnamefont {Maekawa}},
  \bibinfo {author} {\bibfnamefont {K.}~\bibnamefont {Tsuda}}, \ and\ \bibinfo
  {author} {\bibfnamefont {I.}~\bibnamefont {Tanaka}},\ }\href {\doibase
  10.1103/PhysRevB.89.054303} {\bibfield  {journal} {\bibinfo  {journal} {Phys.
  Rev. B}\ }\textbf {\bibinfo {volume} {89}},\ \bibinfo {pages} {054303}
  (\bibinfo {year} {2014})}\BibitemShut {NoStop}%
\bibitem [{\citenamefont {Ghiringhelli}\ \emph {et~al.}(2015)\citenamefont
  {Ghiringhelli}, \citenamefont {Vybiral}, \citenamefont {Levchenko},
  \citenamefont {Draxl},\ and\ \citenamefont
  {Scheffler}}]{ghiringhelli_big_2015}%
  \BibitemOpen
  \bibfield  {author} {\bibinfo {author} {\bibfnamefont {L.~M.}\ \bibnamefont
  {Ghiringhelli}}, \bibinfo {author} {\bibfnamefont {J.}~\bibnamefont
  {Vybiral}}, \bibinfo {author} {\bibfnamefont {S.~V.}\ \bibnamefont
  {Levchenko}}, \bibinfo {author} {\bibfnamefont {C.}~\bibnamefont {Draxl}}, \
  and\ \bibinfo {author} {\bibfnamefont {M.}~\bibnamefont {Scheffler}},\ }\href
  {\doibase 10.1103/PhysRevLett.114.105503} {\bibfield  {journal} {\bibinfo
  {journal} {Phys. Rev. Lett.}\ }\textbf {\bibinfo {volume} {114}},\ \bibinfo
  {pages} {105503} (\bibinfo {year} {2015})}\BibitemShut {NoStop}%
\bibitem [{\citenamefont {Meredig}\ \emph {et~al.}(2014)\citenamefont
  {Meredig}, \citenamefont {Agrawal}, \citenamefont {Kirklin}, \citenamefont
  {Saal}, \citenamefont {Doak}, \citenamefont {Thompson}, \citenamefont
  {Zhang}, \citenamefont {Choudhary},\ and\ \citenamefont
  {Wolverton}}]{meredig_combinatorial_2014}%
  \BibitemOpen
  \bibfield  {author} {\bibinfo {author} {\bibfnamefont {B.}~\bibnamefont
  {Meredig}}, \bibinfo {author} {\bibfnamefont {A.}~\bibnamefont {Agrawal}},
  \bibinfo {author} {\bibfnamefont {S.}~\bibnamefont {Kirklin}}, \bibinfo
  {author} {\bibfnamefont {J.~E.}\ \bibnamefont {Saal}}, \bibinfo {author}
  {\bibfnamefont {J.~W.}\ \bibnamefont {Doak}}, \bibinfo {author}
  {\bibfnamefont {A.}~\bibnamefont {Thompson}}, \bibinfo {author}
  {\bibfnamefont {K.}~\bibnamefont {Zhang}}, \bibinfo {author} {\bibfnamefont
  {A.}~\bibnamefont {Choudhary}}, \ and\ \bibinfo {author} {\bibfnamefont
  {C.}~\bibnamefont {Wolverton}},\ }\href {\doibase 10.1103/PhysRevB.89.094104}
  {\bibfield  {journal} {\bibinfo  {journal} {Phys. Rev. B}\ }\textbf {\bibinfo
  {volume} {89}},\ \bibinfo {pages} {094104} (\bibinfo {year}
  {2014})}\BibitemShut {NoStop}%
\bibitem [{\citenamefont {Deml}\ \emph {et~al.}(2016)\citenamefont {Deml},
  \citenamefont {O’Hayre}, \citenamefont {Wolverton},\ and\ \citenamefont
  {Stevanović}}]{deml_predicting_2016}%
  \BibitemOpen
  \bibfield  {author} {\bibinfo {author} {\bibfnamefont {A.~M.}\ \bibnamefont
  {Deml}}, \bibinfo {author} {\bibfnamefont {R.}~\bibnamefont {O’Hayre}},
  \bibinfo {author} {\bibfnamefont {C.}~\bibnamefont {Wolverton}}, \ and\
  \bibinfo {author} {\bibfnamefont {V.}~\bibnamefont {Stevanović}},\ }\href
  {\doibase 10.1103/PhysRevB.93.085142} {\bibfield  {journal} {\bibinfo
  {journal} {Phys. Rev. B}\ }\textbf {\bibinfo {volume} {93}},\ \bibinfo
  {pages} {085142} (\bibinfo {year} {2016})}\BibitemShut {NoStop}%
\bibitem [{\citenamefont {Bartel}\ \emph {et~al.}(2018)\citenamefont {Bartel},
  \citenamefont {Millican}, \citenamefont {Deml}, \citenamefont {Rumptz},
  \citenamefont {Tumas}, \citenamefont {Weimer}, \citenamefont {Lany},
  \citenamefont {Stevanovi{\'{c}}}, \citenamefont {Musgrave},\ and\
  \citenamefont {Holder}}]{bartel_gibbsML_2018}%
  \BibitemOpen
  \bibfield  {author} {\bibinfo {author} {\bibfnamefont {C.~J.}\ \bibnamefont
  {Bartel}}, \bibinfo {author} {\bibfnamefont {S.~L.}\ \bibnamefont
  {Millican}}, \bibinfo {author} {\bibfnamefont {A.~M.}\ \bibnamefont {Deml}},
  \bibinfo {author} {\bibfnamefont {J.~R.}\ \bibnamefont {Rumptz}}, \bibinfo
  {author} {\bibfnamefont {W.}~\bibnamefont {Tumas}}, \bibinfo {author}
  {\bibfnamefont {A.~W.}\ \bibnamefont {Weimer}}, \bibinfo {author}
  {\bibfnamefont {S.}~\bibnamefont {Lany}}, \bibinfo {author} {\bibfnamefont
  {V.}~\bibnamefont {Stevanovi{\'{c}}}}, \bibinfo {author} {\bibfnamefont
  {C.~B.}\ \bibnamefont {Musgrave}}, \ and\ \bibinfo {author} {\bibfnamefont
  {A.~M.}\ \bibnamefont {Holder}},\ }\href {\doibase
  10.1038/s41467-018-06682-4} {\bibfield  {journal} {\bibinfo  {journal}
  {Nature Communications}\ }\textbf {\bibinfo {volume} {9}},\ \bibinfo {pages}
  {4168} (\bibinfo {year} {2018})},\ \Eprint {http://arxiv.org/abs/1805.08155}
  {1805.08155} \BibitemShut {NoStop}%
\bibitem [{\citenamefont {Isayev}\ \emph {et~al.}(2017)\citenamefont {Isayev},
  \citenamefont {Oses}, \citenamefont {Toher}, \citenamefont {Gossett},
  \citenamefont {Curtarolo},\ and\ \citenamefont
  {Tropsha}}]{isayev_fragment_2017}%
  \BibitemOpen
  \bibfield  {author} {\bibinfo {author} {\bibfnamefont {O.}~\bibnamefont
  {Isayev}}, \bibinfo {author} {\bibfnamefont {C.}~\bibnamefont {Oses}},
  \bibinfo {author} {\bibfnamefont {C.}~\bibnamefont {Toher}}, \bibinfo
  {author} {\bibfnamefont {E.}~\bibnamefont {Gossett}}, \bibinfo {author}
  {\bibfnamefont {S.}~\bibnamefont {Curtarolo}}, \ and\ \bibinfo {author}
  {\bibfnamefont {A.}~\bibnamefont {Tropsha}},\ }\href {\doibase
  10.1038/ncomms15679} {\bibfield  {journal} {\bibinfo  {journal} {Nature
  Communications}\ }\textbf {\bibinfo {volume} {8}},\ \bibinfo {pages} {15679}
  (\bibinfo {year} {2017})},\ \Eprint {http://arxiv.org/abs/1608.04782}
  {1608.04782} \BibitemShut {NoStop}%
\bibitem [{\citenamefont {Faber}\ \emph {et~al.}(2016)\citenamefont {Faber},
  \citenamefont {Lindmaa}, \citenamefont {von Lilienfeld},\ and\ \citenamefont
  {Armiento}}]{faber_machine_2016}%
  \BibitemOpen
  \bibfield  {author} {\bibinfo {author} {\bibfnamefont {F.~A.}\ \bibnamefont
  {Faber}}, \bibinfo {author} {\bibfnamefont {A.}~\bibnamefont {Lindmaa}},
  \bibinfo {author} {\bibfnamefont {O.~A.}\ \bibnamefont {von Lilienfeld}}, \
  and\ \bibinfo {author} {\bibfnamefont {R.}~\bibnamefont {Armiento}},\ }\href
  {\doibase 10.1103/PhysRevLett.117.135502} {\bibfield  {journal} {\bibinfo
  {journal} {Phys. Rev. Lett.}\ }\textbf {\bibinfo {volume} {117}},\ \bibinfo
  {pages} {135502} (\bibinfo {year} {2016})}\BibitemShut {NoStop}%
\bibitem [{\citenamefont {Seko}\ \emph {et~al.}(2018)\citenamefont {Seko},
  \citenamefont {Hayashi}, \citenamefont {Kashima},\ and\ \citenamefont
  {Tanaka}}]{seko_matrix-_2018}%
  \BibitemOpen
  \bibfield  {author} {\bibinfo {author} {\bibfnamefont {A.}~\bibnamefont
  {Seko}}, \bibinfo {author} {\bibfnamefont {H.}~\bibnamefont {Hayashi}},
  \bibinfo {author} {\bibfnamefont {H.}~\bibnamefont {Kashima}}, \ and\
  \bibinfo {author} {\bibfnamefont {I.}~\bibnamefont {Tanaka}},\ }\href
  {\doibase 10.1103/PhysRevMaterials.2.013805} {\bibfield  {journal} {\bibinfo
  {journal} {Phys. Rev. Materials}\ }\textbf {\bibinfo {volume} {2}},\ \bibinfo
  {pages} {013805} (\bibinfo {year} {2018})}\BibitemShut {NoStop}%
\bibitem [{\citenamefont {Schmidt}\ \emph {et~al.}(2017)\citenamefont
  {Schmidt}, \citenamefont {Shi}, \citenamefont {Borlido}, \citenamefont
  {Chen}, \citenamefont {Botti},\ and\ \citenamefont
  {Marques}}]{schmidt_thermo_2017}%
  \BibitemOpen
  \bibfield  {author} {\bibinfo {author} {\bibfnamefont {J.}~\bibnamefont
  {Schmidt}}, \bibinfo {author} {\bibfnamefont {J.}~\bibnamefont {Shi}},
  \bibinfo {author} {\bibfnamefont {P.}~\bibnamefont {Borlido}}, \bibinfo
  {author} {\bibfnamefont {L.}~\bibnamefont {Chen}}, \bibinfo {author}
  {\bibfnamefont {S.}~\bibnamefont {Botti}}, \ and\ \bibinfo {author}
  {\bibfnamefont {M.~A.~L.}\ \bibnamefont {Marques}},\ }\href {\doibase
  10.1021/acs.chemmater.7b00156} {\bibfield  {journal} {\bibinfo  {journal}
  {Chemistry of Materials}\ }\textbf {\bibinfo {volume} {29}},\ \bibinfo
  {pages} {5090} (\bibinfo {year} {2017})},\ \Eprint
  {http://arxiv.org/abs/arXiv:1310.1546v1} {arXiv:arXiv:1310.1546v1}
  \BibitemShut {NoStop}%
\bibitem [{\citenamefont {Bhadeshia}\ \emph {et~al.}(2009)\citenamefont
  {Bhadeshia}, \citenamefont {Dimitriu}, \citenamefont {Forsik}, \citenamefont
  {Pak},\ and\ \citenamefont {Ryu}}]{bhadeshia_performance_2009}%
  \BibitemOpen
  \bibfield  {author} {\bibinfo {author} {\bibfnamefont {H.~K. D.~H.}\
  \bibnamefont {Bhadeshia}}, \bibinfo {author} {\bibfnamefont {R.~C.}\
  \bibnamefont {Dimitriu}}, \bibinfo {author} {\bibfnamefont {S.}~\bibnamefont
  {Forsik}}, \bibinfo {author} {\bibfnamefont {J.~H.}\ \bibnamefont {Pak}}, \
  and\ \bibinfo {author} {\bibfnamefont {J.~H.}\ \bibnamefont {Ryu}},\ }\href
  {\doibase 10.1179/174328408X311053} {\bibfield  {journal} {\bibinfo
  {journal} {Materials Science and Technology}\ }\textbf {\bibinfo {volume}
  {25}},\ \bibinfo {pages} {504} (\bibinfo {year} {2009})}\BibitemShut
  {NoStop}%
\bibitem [{\citenamefont {Chatterjee}\ \emph {et~al.}(2007)\citenamefont
  {Chatterjee}, \citenamefont {Murugananth},\ and\ \citenamefont
  {Bhadeshia}}]{chatterjee__2007}%
  \BibitemOpen
  \bibfield  {author} {\bibinfo {author} {\bibfnamefont {S.}~\bibnamefont
  {Chatterjee}}, \bibinfo {author} {\bibfnamefont {M.}~\bibnamefont
  {Murugananth}}, \ and\ \bibinfo {author} {\bibfnamefont {H.~K. D.~H.}\
  \bibnamefont {Bhadeshia}},\ }\href {\doibase 10.1179/174328407X179746}
  {\bibfield  {journal} {\bibinfo  {journal} {Materials Science and
  Technology}\ }\textbf {\bibinfo {volume} {23}},\ \bibinfo {pages} {819}
  (\bibinfo {year} {2007})}\BibitemShut {NoStop}%
\bibitem [{\citenamefont {Xue}\ \emph {et~al.}(2016)\citenamefont {Xue},
  \citenamefont {Balachandran}, \citenamefont {Hogden}, \citenamefont
  {Theiler}, \citenamefont {Xue},\ and\ \citenamefont
  {Lookman}}]{xue_sma_2016}%
  \BibitemOpen
  \bibfield  {author} {\bibinfo {author} {\bibfnamefont {D.}~\bibnamefont
  {Xue}}, \bibinfo {author} {\bibfnamefont {P.~V.}\ \bibnamefont
  {Balachandran}}, \bibinfo {author} {\bibfnamefont {J.}~\bibnamefont
  {Hogden}}, \bibinfo {author} {\bibfnamefont {J.}~\bibnamefont {Theiler}},
  \bibinfo {author} {\bibfnamefont {D.}~\bibnamefont {Xue}}, \ and\ \bibinfo
  {author} {\bibfnamefont {T.}~\bibnamefont {Lookman}},\ }\href {\doibase
  10.1038/ncomms11241} {\bibfield  {journal} {\bibinfo  {journal} {Nature
  Communications}\ }\textbf {\bibinfo {volume} {7}},\ \bibinfo {pages} {11241}
  (\bibinfo {year} {2016})}\BibitemShut {NoStop}%
\bibitem [{\citenamefont {Stanev}\ \emph {et~al.}(2018)\citenamefont {Stanev},
  \citenamefont {Oses}, \citenamefont {Kusne}, \citenamefont {Rodriguez},
  \citenamefont {Paglione}, \citenamefont {Curtarolo},\ and\ \citenamefont
  {Takeuchi}}]{stanev_machine_2018}%
  \BibitemOpen
  \bibfield  {author} {\bibinfo {author} {\bibfnamefont {V.}~\bibnamefont
  {Stanev}}, \bibinfo {author} {\bibfnamefont {C.}~\bibnamefont {Oses}},
  \bibinfo {author} {\bibfnamefont {A.~G.}\ \bibnamefont {Kusne}}, \bibinfo
  {author} {\bibfnamefont {E.}~\bibnamefont {Rodriguez}}, \bibinfo {author}
  {\bibfnamefont {J.}~\bibnamefont {Paglione}}, \bibinfo {author}
  {\bibfnamefont {S.}~\bibnamefont {Curtarolo}}, \ and\ \bibinfo {author}
  {\bibfnamefont {I.}~\bibnamefont {Takeuchi}},\ }\href {\doibase
  10.1038/s41524-018-0085-8} {\bibfield  {journal} {\bibinfo  {journal} {npj
  Computational Materials}\ }\textbf {\bibinfo {volume} {4}},\ \bibinfo {pages}
  {29} (\bibinfo {year} {2018})}\BibitemShut {NoStop}%
\bibitem [{\citenamefont {Ward}\ \emph {et~al.}(2018)\citenamefont {Ward},
  \citenamefont {O'Keeffe}, \citenamefont {Stevick}, \citenamefont {Jelbert},
  \citenamefont {Aykol},\ and\ \citenamefont {Wolverton}}]{ward_machine_2018}%
  \BibitemOpen
  \bibfield  {author} {\bibinfo {author} {\bibfnamefont {L.}~\bibnamefont
  {Ward}}, \bibinfo {author} {\bibfnamefont {S.~C.}\ \bibnamefont {O'Keeffe}},
  \bibinfo {author} {\bibfnamefont {J.}~\bibnamefont {Stevick}}, \bibinfo
  {author} {\bibfnamefont {G.~R.}\ \bibnamefont {Jelbert}}, \bibinfo {author}
  {\bibfnamefont {M.}~\bibnamefont {Aykol}}, \ and\ \bibinfo {author}
  {\bibfnamefont {C.}~\bibnamefont {Wolverton}},\ }\href {\doibase
  10.1016/j.actamat.2018.08.002} {\bibfield  {journal} {\bibinfo  {journal}
  {Acta Materialia}\ }\textbf {\bibinfo {volume} {159}},\ \bibinfo {pages}
  {102} (\bibinfo {year} {2018})}\BibitemShut {NoStop}%
\bibitem [{\citenamefont {Dey}\ \emph {et~al.}(2014)\citenamefont {Dey},
  \citenamefont {Bible}, \citenamefont {Datta}, \citenamefont {Broderick},
  \citenamefont {Jasinski}, \citenamefont {Sunkara}, \citenamefont {Menon},\
  and\ \citenamefont {Rajan}}]{dey_informatics-aided_2014}%
  \BibitemOpen
  \bibfield  {author} {\bibinfo {author} {\bibfnamefont {P.}~\bibnamefont
  {Dey}}, \bibinfo {author} {\bibfnamefont {J.}~\bibnamefont {Bible}}, \bibinfo
  {author} {\bibfnamefont {S.}~\bibnamefont {Datta}}, \bibinfo {author}
  {\bibfnamefont {S.}~\bibnamefont {Broderick}}, \bibinfo {author}
  {\bibfnamefont {J.}~\bibnamefont {Jasinski}}, \bibinfo {author}
  {\bibfnamefont {M.}~\bibnamefont {Sunkara}}, \bibinfo {author} {\bibfnamefont
  {M.}~\bibnamefont {Menon}}, \ and\ \bibinfo {author} {\bibfnamefont
  {K.}~\bibnamefont {Rajan}},\ }\href {\doibase
  10.1016/j.commatsci.2013.10.016} {\bibfield  {journal} {\bibinfo  {journal}
  {Computational Materials Science}\ }\textbf {\bibinfo {volume} {83}},\
  \bibinfo {pages} {185} (\bibinfo {year} {2014})}\BibitemShut {NoStop}%
\bibitem [{\citenamefont {Pilania}\ \emph {et~al.}(2016)\citenamefont
  {Pilania}, \citenamefont {Mannodi-Kanakkithodi}, \citenamefont {Uberuaga},
  \citenamefont {Ramprasad}, \citenamefont {Gubernatis},\ and\ \citenamefont
  {Lookman}}]{pilania_machine_2016}%
  \BibitemOpen
  \bibfield  {author} {\bibinfo {author} {\bibfnamefont {G.}~\bibnamefont
  {Pilania}}, \bibinfo {author} {\bibfnamefont {A.}~\bibnamefont
  {Mannodi-Kanakkithodi}}, \bibinfo {author} {\bibfnamefont {B.~P.}\
  \bibnamefont {Uberuaga}}, \bibinfo {author} {\bibfnamefont {R.}~\bibnamefont
  {Ramprasad}}, \bibinfo {author} {\bibfnamefont {J.~E.}\ \bibnamefont
  {Gubernatis}}, \ and\ \bibinfo {author} {\bibfnamefont {T.}~\bibnamefont
  {Lookman}},\ }\href {\doibase 10.1038/srep19375} {\bibfield  {journal}
  {\bibinfo  {journal} {Scientific Reports}\ }\textbf {\bibinfo {volume} {6}},\
  \bibinfo {pages} {19375} (\bibinfo {year} {2016})}\BibitemShut {NoStop}%
\bibitem [{\citenamefont {Ward}\ \emph {et~al.}(2016)\citenamefont {Ward},
  \citenamefont {Agrawal}, \citenamefont {Choudhary},\ and\ \citenamefont
  {Wolverton}}]{ward_general-purpose_2016}%
  \BibitemOpen
  \bibfield  {author} {\bibinfo {author} {\bibfnamefont {L.}~\bibnamefont
  {Ward}}, \bibinfo {author} {\bibfnamefont {A.}~\bibnamefont {Agrawal}},
  \bibinfo {author} {\bibfnamefont {A.}~\bibnamefont {Choudhary}}, \ and\
  \bibinfo {author} {\bibfnamefont {C.}~\bibnamefont {Wolverton}},\ }\href
  {\doibase 10.1038/npjcompumats.2016.28} {\bibfield  {journal} {\bibinfo
  {journal} {npj Computational Materials}\ }\textbf {\bibinfo {volume} {2}},\
  \bibinfo {pages} {16028} (\bibinfo {year} {2016})}\BibitemShut {NoStop}%
\bibitem [{\citenamefont {Huan}\ \emph {et~al.}(2015)\citenamefont {Huan},
  \citenamefont {Mannodi-Kanakkithodi},\ and\ \citenamefont
  {Ramprasad}}]{huan_motif_2015}%
  \BibitemOpen
  \bibfield  {author} {\bibinfo {author} {\bibfnamefont {T.~D.}\ \bibnamefont
  {Huan}}, \bibinfo {author} {\bibfnamefont {A.}~\bibnamefont
  {Mannodi-Kanakkithodi}}, \ and\ \bibinfo {author} {\bibfnamefont
  {R.}~\bibnamefont {Ramprasad}},\ }\href {\doibase 10.1103/PhysRevB.92.014106}
  {\bibfield  {journal} {\bibinfo  {journal} {Physical Review B}\ }\textbf
  {\bibinfo {volume} {92}},\ \bibinfo {pages} {014106} (\bibinfo {year}
  {2015})}\BibitemShut {NoStop}%
\bibitem [{\citenamefont {Sparks}\ \emph {et~al.}(2015)\citenamefont {Sparks},
  \citenamefont {Gaultois}, \citenamefont {Oliynyk}, \citenamefont {Brgoch},\
  and\ \citenamefont {Meredig}}]{sparks_te_2015}%
  \BibitemOpen
  \bibfield  {author} {\bibinfo {author} {\bibfnamefont {T.~D.}\ \bibnamefont
  {Sparks}}, \bibinfo {author} {\bibfnamefont {M.~W.}\ \bibnamefont
  {Gaultois}}, \bibinfo {author} {\bibfnamefont {A.}~\bibnamefont {Oliynyk}},
  \bibinfo {author} {\bibfnamefont {J.}~\bibnamefont {Brgoch}}, \ and\ \bibinfo
  {author} {\bibfnamefont {B.}~\bibnamefont {Meredig}},\ }\href {\doibase
  10.1016/j.scriptamat.2015.04.026} {\bibfield  {journal} {\bibinfo  {journal}
  {Scripta Materialia}\ }\textbf {\bibinfo {volume} {111}},\ \bibinfo {pages}
  {10} (\bibinfo {year} {2015})}\BibitemShut {NoStop}%
\bibitem [{\citenamefont {Goedecker}(2004)}]{goedecker_minima_2004}%
  \BibitemOpen
  \bibfield  {author} {\bibinfo {author} {\bibfnamefont {S.}~\bibnamefont
  {Goedecker}},\ }\href@noop {} {\bibfield  {journal} {\bibinfo  {journal} {J.
  Chem. Phys.}\ }\textbf {\bibinfo {volume} {120}},\ \bibinfo {pages} {9911}
  (\bibinfo {year} {2004})}\BibitemShut {NoStop}%
\bibitem [{\citenamefont {Amsler}\ and\ \citenamefont
  {Goedecker}(2010)}]{amsler_crystal_2010}%
  \BibitemOpen
  \bibfield  {author} {\bibinfo {author} {\bibfnamefont {M.}~\bibnamefont
  {Amsler}}\ and\ \bibinfo {author} {\bibfnamefont {S.}~\bibnamefont
  {Goedecker}},\ }\href@noop {} {\bibfield  {journal} {\bibinfo  {journal} {J.
  Chem. Phys.}\ }\textbf {\bibinfo {volume} {133}},\ \bibinfo {pages} {224104}
  (\bibinfo {year} {2010})}\BibitemShut {NoStop}%
\bibitem [{\citenamefont {Zhao}\ \emph {et~al.}(2014)\citenamefont {Zhao},
  \citenamefont {Lo}, \citenamefont {Zhang}, \citenamefont {Sun}, \citenamefont
  {Tan}, \citenamefont {Uher}, \citenamefont {Wolverton}, \citenamefont
  {Dravid},\ and\ \citenamefont {Kanatzidis}}]{SnSe}%
  \BibitemOpen
  \bibfield  {author} {\bibinfo {author} {\bibfnamefont {L.-D.}\ \bibnamefont
  {Zhao}}, \bibinfo {author} {\bibfnamefont {S.-H.}\ \bibnamefont {Lo}},
  \bibinfo {author} {\bibfnamefont {Y.}~\bibnamefont {Zhang}}, \bibinfo
  {author} {\bibfnamefont {H.}~\bibnamefont {Sun}}, \bibinfo {author}
  {\bibfnamefont {G.}~\bibnamefont {Tan}}, \bibinfo {author} {\bibfnamefont
  {C.}~\bibnamefont {Uher}}, \bibinfo {author} {\bibfnamefont {C.}~\bibnamefont
  {Wolverton}}, \bibinfo {author} {\bibfnamefont {V.~P.}\ \bibnamefont
  {Dravid}}, \ and\ \bibinfo {author} {\bibfnamefont {M.~G.}\ \bibnamefont
  {Kanatzidis}},\ }\href {\doibase 10.1038/nature13184} {\bibfield  {journal}
  {\bibinfo  {journal} {Nature}\ }\textbf {\bibinfo {volume} {508}},\ \bibinfo
  {pages} {373} (\bibinfo {year} {2014})}\BibitemShut {NoStop}%
\bibitem [{\citenamefont {Li}\ \emph {et~al.}(2015)\citenamefont {Li},
  \citenamefont {Hong}, \citenamefont {May}, \citenamefont {Bansal},
  \citenamefont {Chi}, \citenamefont {Hong}, \citenamefont {Ehlers},\ and\
  \citenamefont {Delaire}}]{doi:10.1038/nphys3492}%
  \BibitemOpen
  \bibfield  {author} {\bibinfo {author} {\bibfnamefont {C.~W.}\ \bibnamefont
  {Li}}, \bibinfo {author} {\bibfnamefont {J.}~\bibnamefont {Hong}}, \bibinfo
  {author} {\bibfnamefont {A.~F.}\ \bibnamefont {May}}, \bibinfo {author}
  {\bibfnamefont {D.}~\bibnamefont {Bansal}}, \bibinfo {author} {\bibfnamefont
  {S.}~\bibnamefont {Chi}}, \bibinfo {author} {\bibfnamefont {T.}~\bibnamefont
  {Hong}}, \bibinfo {author} {\bibfnamefont {G.}~\bibnamefont {Ehlers}}, \ and\
  \bibinfo {author} {\bibfnamefont {O.}~\bibnamefont {Delaire}},\ }\href
  {\doibase 10.1038/nphys3492} {\bibfield  {journal} {\bibinfo  {journal} {Nat.
  Phys.}\ }\textbf {\bibinfo {volume} {11}},\ \bibinfo {pages} {1063} (\bibinfo
  {year} {2015})}\BibitemShut {NoStop}%
\bibitem [{\citenamefont {Breiman}(2001)}]{breiman_randomforest_2001}%
  \BibitemOpen
  \bibfield  {author} {\bibinfo {author} {\bibfnamefont {L.}~\bibnamefont
  {Breiman}},\ }\href {\doibase 10.1023/A:1010933404324} {\bibfield  {journal}
  {\bibinfo  {journal} {Machine Learning}\ }\textbf {\bibinfo {volume} {45}},\
  \bibinfo {pages} {5} (\bibinfo {year} {2001})}\BibitemShut {NoStop}%
\bibitem [{mhm()}]{mhm_github}%
  \BibitemOpen
  \href@noop {} {}\bibinfo {note}
  {\url{https://github.com/WardLT/ternary-semiconductors-mhm}}\BibitemShut
  {NoStop}%
\bibitem [{\citenamefont {Roy}\ \emph {et~al.}(2008)\citenamefont {Roy},
  \citenamefont {Goedecker},\ and\ \citenamefont
  {Hellmann}}]{roy_bell-evans-polanyi_2008}%
  \BibitemOpen
  \bibfield  {author} {\bibinfo {author} {\bibfnamefont {S.}~\bibnamefont
  {Roy}}, \bibinfo {author} {\bibfnamefont {S.}~\bibnamefont {Goedecker}}, \
  and\ \bibinfo {author} {\bibfnamefont {V.}~\bibnamefont {Hellmann}},\ }\href
  {\doibase 10.1103/PhysRevE.77.056707} {\bibfield  {journal} {\bibinfo
  {journal} {Phys. Rev. E}\ }\textbf {\bibinfo {volume} {77}},\ \bibinfo
  {pages} {056707} (\bibinfo {year} {2008})}\BibitemShut {NoStop}%
\bibitem [{\citenamefont {Sicher}\ \emph {et~al.}(2011)\citenamefont {Sicher},
  \citenamefont {Mohr},\ and\ \citenamefont
  {Goedecker}}]{sicher_efficient_2011}%
  \BibitemOpen
  \bibfield  {author} {\bibinfo {author} {\bibfnamefont {M.}~\bibnamefont
  {Sicher}}, \bibinfo {author} {\bibfnamefont {S.}~\bibnamefont {Mohr}}, \ and\
  \bibinfo {author} {\bibfnamefont {S.}~\bibnamefont {Goedecker}},\ }\href
  {\doibase doi:10.1063/1.3530590} {\bibfield  {journal} {\bibinfo  {journal}
  {J. Chem. Phys.}\ }\textbf {\bibinfo {volume} {134}},\ \bibinfo {pages}
  {044106} (\bibinfo {year} {2011})}\BibitemShut {NoStop}%
\bibitem [{\citenamefont {Amsler}\ \emph
  {et~al.}(2012{\natexlab{a}})\citenamefont {Amsler}, \citenamefont
  {{Flores-Livas}}, \citenamefont {Lehtovaara}, \citenamefont {Balima},
  \citenamefont {Ghasemi}, \citenamefont {Machon}, \citenamefont {Pailh\`{e}s},
  \citenamefont {Willand}, \citenamefont {Caliste}, \citenamefont {Botti},
  \citenamefont {San~Miguel}, \citenamefont {Goedecker},\ and\ \citenamefont
  {Marques}}]{amsler_crystal_2012}%
  \BibitemOpen
  \bibfield  {author} {\bibinfo {author} {\bibfnamefont {M.}~\bibnamefont
  {Amsler}}, \bibinfo {author} {\bibfnamefont {J.~A.}\ \bibnamefont
  {{Flores-Livas}}}, \bibinfo {author} {\bibfnamefont {L.}~\bibnamefont
  {Lehtovaara}}, \bibinfo {author} {\bibfnamefont {F.}~\bibnamefont {Balima}},
  \bibinfo {author} {\bibfnamefont {S.~A.}\ \bibnamefont {Ghasemi}}, \bibinfo
  {author} {\bibfnamefont {D.}~\bibnamefont {Machon}}, \bibinfo {author}
  {\bibfnamefont {S.}~\bibnamefont {Pailh\`{e}s}}, \bibinfo {author}
  {\bibfnamefont {A.}~\bibnamefont {Willand}}, \bibinfo {author} {\bibfnamefont
  {D.}~\bibnamefont {Caliste}}, \bibinfo {author} {\bibfnamefont
  {S.}~\bibnamefont {Botti}}, \bibinfo {author} {\bibfnamefont
  {A.}~\bibnamefont {San~Miguel}}, \bibinfo {author} {\bibfnamefont
  {S.}~\bibnamefont {Goedecker}}, \ and\ \bibinfo {author} {\bibfnamefont
  {M.~A.~L.}\ \bibnamefont {Marques}},\ }\href {\doibase
  10.1103/PhysRevLett.108.065501} {\bibfield  {journal} {\bibinfo  {journal}
  {Phys. Rev. Lett.}\ }\textbf {\bibinfo {volume} {108}},\ \bibinfo {pages}
  {065501} (\bibinfo {year} {2012}{\natexlab{a}})}\BibitemShut {NoStop}%
\bibitem [{\citenamefont {{Flores-Livas}}\ \emph {et~al.}(2012)\citenamefont
  {{Flores-Livas}}, \citenamefont {Amsler}, \citenamefont {Lenosky},
  \citenamefont {Lehtovaara}, \citenamefont {Botti}, \citenamefont {Marques},\
  and\ \citenamefont {Goedecker}}]{flores-livas_high-pressure_2012}%
  \BibitemOpen
  \bibfield  {author} {\bibinfo {author} {\bibfnamefont {J.~A.}\ \bibnamefont
  {{Flores-Livas}}}, \bibinfo {author} {\bibfnamefont {M.}~\bibnamefont
  {Amsler}}, \bibinfo {author} {\bibfnamefont {T.~J.}\ \bibnamefont {Lenosky}},
  \bibinfo {author} {\bibfnamefont {L.}~\bibnamefont {Lehtovaara}}, \bibinfo
  {author} {\bibfnamefont {S.}~\bibnamefont {Botti}}, \bibinfo {author}
  {\bibfnamefont {M.~A.~L.}\ \bibnamefont {Marques}}, \ and\ \bibinfo {author}
  {\bibfnamefont {S.}~\bibnamefont {Goedecker}},\ }\href {\doibase
  10.1103/PhysRevLett.108.117004} {\bibfield  {journal} {\bibinfo  {journal}
  {Phys. Rev. Lett.}\ }\textbf {\bibinfo {volume} {108}},\ \bibinfo {pages}
  {117004} (\bibinfo {year} {2012})}\BibitemShut {NoStop}%
\bibitem [{\citenamefont {Amsler}\ \emph
  {et~al.}(2012{\natexlab{b}})\citenamefont {Amsler}, \citenamefont
  {Flores-Livas}, \citenamefont {Huan}, \citenamefont {Botti}, \citenamefont
  {Marques},\ and\ \citenamefont {Goedecker}}]{amsler_novel_2012}%
  \BibitemOpen
  \bibfield  {author} {\bibinfo {author} {\bibfnamefont {M.}~\bibnamefont
  {Amsler}}, \bibinfo {author} {\bibfnamefont {J.~A.}\ \bibnamefont
  {Flores-Livas}}, \bibinfo {author} {\bibfnamefont {T.~D.}\ \bibnamefont
  {Huan}}, \bibinfo {author} {\bibfnamefont {S.}~\bibnamefont {Botti}},
  \bibinfo {author} {\bibfnamefont {M.~A.~L.}\ \bibnamefont {Marques}}, \ and\
  \bibinfo {author} {\bibfnamefont {S.}~\bibnamefont {Goedecker}},\ }\href
  {\doibase 10.1103/PhysRevLett.108.205505} {\bibfield  {journal} {\bibinfo
  {journal} {Phys. Rev. Lett.}\ }\textbf {\bibinfo {volume} {108}},\ \bibinfo
  {pages} {205505} (\bibinfo {year} {2012}{\natexlab{b}})}\BibitemShut
  {NoStop}%
\bibitem [{\citenamefont {Huan}\ \emph {et~al.}(2013)\citenamefont {Huan},
  \citenamefont {Amsler}, \citenamefont {Sabatini}, \citenamefont {Tuoc},
  \citenamefont {Le}, \citenamefont {Woods}, \citenamefont {Marzari},\ and\
  \citenamefont {Goedecker}}]{huan_thermodynamic_2013}%
  \BibitemOpen
  \bibfield  {author} {\bibinfo {author} {\bibfnamefont {T.~D.}\ \bibnamefont
  {Huan}}, \bibinfo {author} {\bibfnamefont {M.}~\bibnamefont {Amsler}},
  \bibinfo {author} {\bibfnamefont {R.}~\bibnamefont {Sabatini}}, \bibinfo
  {author} {\bibfnamefont {V.~N.}\ \bibnamefont {Tuoc}}, \bibinfo {author}
  {\bibfnamefont {N.~B.}\ \bibnamefont {Le}}, \bibinfo {author} {\bibfnamefont
  {L.~M.}\ \bibnamefont {Woods}}, \bibinfo {author} {\bibfnamefont
  {N.}~\bibnamefont {Marzari}}, \ and\ \bibinfo {author} {\bibfnamefont
  {S.}~\bibnamefont {Goedecker}},\ }\href {\doibase 10.1103/PhysRevB.88.024108}
  {\bibfield  {journal} {\bibinfo  {journal} {Phys. Rev. B}\ }\textbf {\bibinfo
  {volume} {88}},\ \bibinfo {pages} {024108} (\bibinfo {year}
  {2013})}\BibitemShut {NoStop}%
\bibitem [{\citenamefont {Clarke}\ \emph {et~al.}(2016)\citenamefont {Clarke},
  \citenamefont {Walsh}, \citenamefont {Amsler}, \citenamefont {Malliakas},
  \citenamefont {Yu}, \citenamefont {Goedecker}, \citenamefont {Wang},
  \citenamefont {Wolverton},\ and\ \citenamefont
  {Freedman}}]{clarke_discovery_2016}%
  \BibitemOpen
  \bibfield  {author} {\bibinfo {author} {\bibfnamefont {S.~M.}\ \bibnamefont
  {Clarke}}, \bibinfo {author} {\bibfnamefont {J.~P.~S.}\ \bibnamefont
  {Walsh}}, \bibinfo {author} {\bibfnamefont {M.}~\bibnamefont {Amsler}},
  \bibinfo {author} {\bibfnamefont {C.~D.}\ \bibnamefont {Malliakas}}, \bibinfo
  {author} {\bibfnamefont {T.}~\bibnamefont {Yu}}, \bibinfo {author}
  {\bibfnamefont {S.}~\bibnamefont {Goedecker}}, \bibinfo {author}
  {\bibfnamefont {Y.}~\bibnamefont {Wang}}, \bibinfo {author} {\bibfnamefont
  {C.}~\bibnamefont {Wolverton}}, \ and\ \bibinfo {author} {\bibfnamefont
  {D.~E.}\ \bibnamefont {Freedman}},\ }\href {\doibase 10.1002/anie.201605902}
  {\bibfield  {journal} {\bibinfo  {journal} {Angew. Chem. Int. Ed.}\ }\textbf
  {\bibinfo {volume} {55}},\ \bibinfo {pages} {13446} (\bibinfo {year}
  {2016})}\BibitemShut {NoStop}%
\bibitem [{\citenamefont {Clarke}\ \emph {et~al.}(2017)\citenamefont {Clarke},
  \citenamefont {Amsler}, \citenamefont {Walsh}, \citenamefont {Yu},
  \citenamefont {Wang}, \citenamefont {Meng}, \citenamefont {Jacobsen},
  \citenamefont {Wolverton},\ and\ \citenamefont
  {Freedman}}]{clarke_creating_2017}%
  \BibitemOpen
  \bibfield  {author} {\bibinfo {author} {\bibfnamefont {S.~M.}\ \bibnamefont
  {Clarke}}, \bibinfo {author} {\bibfnamefont {M.}~\bibnamefont {Amsler}},
  \bibinfo {author} {\bibfnamefont {J.~P.~S.}\ \bibnamefont {Walsh}}, \bibinfo
  {author} {\bibfnamefont {T.}~\bibnamefont {Yu}}, \bibinfo {author}
  {\bibfnamefont {Y.}~\bibnamefont {Wang}}, \bibinfo {author} {\bibfnamefont
  {Y.}~\bibnamefont {Meng}}, \bibinfo {author} {\bibfnamefont {S.~D.}\
  \bibnamefont {Jacobsen}}, \bibinfo {author} {\bibfnamefont {C.}~\bibnamefont
  {Wolverton}}, \ and\ \bibinfo {author} {\bibfnamefont {D.~E.}\ \bibnamefont
  {Freedman}},\ }\href {\doibase 10.1021/acs.chemmater.7b01418} {\bibfield
  {journal} {\bibinfo  {journal} {Chem. Mater.}\ }\textbf {\bibinfo {volume}
  {29}},\ \bibinfo {pages} {5276} (\bibinfo {year} {2017})}\BibitemShut
  {NoStop}%
\bibitem [{\citenamefont {Amsler}\ \emph {et~al.}(2017)\citenamefont {Amsler},
  \citenamefont {Naghavi},\ and\ \citenamefont
  {Wolverton}}]{amsler_prediction_2017}%
  \BibitemOpen
  \bibfield  {author} {\bibinfo {author} {\bibfnamefont {M.}~\bibnamefont
  {Amsler}}, \bibinfo {author} {\bibfnamefont {S.~S.}\ \bibnamefont {Naghavi}},
  \ and\ \bibinfo {author} {\bibfnamefont {C.}~\bibnamefont {Wolverton}},\
  }\href {\doibase 10.1039/C6SC04683E} {\bibfield  {journal} {\bibinfo
  {journal} {Chem. Sci.}\ }\textbf {\bibinfo {volume} {8}},\ \bibinfo {pages}
  {2226} (\bibinfo {year} {2017})}\BibitemShut {NoStop}%
\bibitem [{\citenamefont {Amsler}\ \emph {et~al.}(2018)\citenamefont {Amsler},
  \citenamefont {Hegde}, \citenamefont {Jacobsen},\ and\ \citenamefont
  {Wolverton}}]{amsler_exploring_2018}%
  \BibitemOpen
  \bibfield  {author} {\bibinfo {author} {\bibfnamefont {M.}~\bibnamefont
  {Amsler}}, \bibinfo {author} {\bibfnamefont {V.~I.}\ \bibnamefont {Hegde}},
  \bibinfo {author} {\bibfnamefont {S.~D.}\ \bibnamefont {Jacobsen}}, \ and\
  \bibinfo {author} {\bibfnamefont {C.}~\bibnamefont {Wolverton}},\ }\href
  {\doibase 10.1103/PhysRevX.8.041021} {\bibfield  {journal} {\bibinfo
  {journal} {Phys. Rev. X}\ }\textbf {\bibinfo {volume} {8}},\ \bibinfo {pages}
  {041021} (\bibinfo {year} {2018})}\BibitemShut {NoStop}%
\bibitem [{\citenamefont {Kresse}\ and\ \citenamefont
  {Hafner}(1993)}]{kresse1993ab}%
  \BibitemOpen
  \bibfield  {author} {\bibinfo {author} {\bibfnamefont {G.}~\bibnamefont
  {Kresse}}\ and\ \bibinfo {author} {\bibfnamefont {J.}~\bibnamefont
  {Hafner}},\ }\href@noop {} {\bibfield  {journal} {\bibinfo  {journal} {Phys.
  Rev. B}\ }\textbf {\bibinfo {volume} {47}},\ \bibinfo {pages} {558} (\bibinfo
  {year} {1993})}\BibitemShut {NoStop}%
\bibitem [{\citenamefont {Kresse}\ and\ \citenamefont
  {Furthm{\"u}ller}(1996{\natexlab{a}})}]{kresse1996efficiency}%
  \BibitemOpen
  \bibfield  {author} {\bibinfo {author} {\bibfnamefont {G.}~\bibnamefont
  {Kresse}}\ and\ \bibinfo {author} {\bibfnamefont {J.}~\bibnamefont
  {Furthm{\"u}ller}},\ }\href@noop {} {\bibfield  {journal} {\bibinfo
  {journal} {Comput. Mater. Sci.}\ }\textbf {\bibinfo {volume} {6}},\ \bibinfo
  {pages} {15} (\bibinfo {year} {1996}{\natexlab{a}})}\BibitemShut {NoStop}%
\bibitem [{\citenamefont {Kresse}\ and\ \citenamefont
  {Furthm{\"u}ller}(1996{\natexlab{b}})}]{kresse1996efficient}%
  \BibitemOpen
  \bibfield  {author} {\bibinfo {author} {\bibfnamefont {G.}~\bibnamefont
  {Kresse}}\ and\ \bibinfo {author} {\bibfnamefont {J.}~\bibnamefont
  {Furthm{\"u}ller}},\ }\href@noop {} {\bibfield  {journal} {\bibinfo
  {journal} {Phys. Rev. B}\ }\textbf {\bibinfo {volume} {54}},\ \bibinfo
  {pages} {11169} (\bibinfo {year} {1996}{\natexlab{b}})}\BibitemShut {NoStop}%
\bibitem [{\citenamefont {Bl{\"o}chl}(1994)}]{blochl1994projector}%
  \BibitemOpen
  \bibfield  {author} {\bibinfo {author} {\bibfnamefont {P.~E.}\ \bibnamefont
  {Bl{\"o}chl}},\ }\href@noop {} {\bibfield  {journal} {\bibinfo  {journal}
  {Phys. Rev. B}\ }\textbf {\bibinfo {volume} {50}},\ \bibinfo {pages} {17953}
  (\bibinfo {year} {1994})}\BibitemShut {NoStop}%
\bibitem [{\citenamefont {Kresse}\ and\ \citenamefont
  {Joubert}(1999)}]{kresse_paw_1999}%
  \BibitemOpen
  \bibfield  {author} {\bibinfo {author} {\bibfnamefont {G.}~\bibnamefont
  {Kresse}}\ and\ \bibinfo {author} {\bibfnamefont {D.}~\bibnamefont
  {Joubert}},\ }\href {\doibase 10.1103/PhysRevB.59.1758} {\bibfield  {journal}
  {\bibinfo  {journal} {Phys. Rev. B}\ }\textbf {\bibinfo {volume} {59}},\
  \bibinfo {pages} {1758} (\bibinfo {year} {1999})}\BibitemShut {NoStop}%
\bibitem [{\citenamefont {Perdew}\ \emph {et~al.}(1996)\citenamefont {Perdew},
  \citenamefont {Burke},\ and\ \citenamefont
  {Ernzerhof}}]{perdew1996generalized}%
  \BibitemOpen
  \bibfield  {author} {\bibinfo {author} {\bibfnamefont {J.~P.}\ \bibnamefont
  {Perdew}}, \bibinfo {author} {\bibfnamefont {K.}~\bibnamefont {Burke}}, \
  and\ \bibinfo {author} {\bibfnamefont {M.}~\bibnamefont {Ernzerhof}},\
  }\href@noop {} {\bibfield  {journal} {\bibinfo  {journal} {Phys. Rev. Lett.}\
  }\textbf {\bibinfo {volume} {77}},\ \bibinfo {pages} {3865} (\bibinfo {year}
  {1996})}\BibitemShut {NoStop}%
\bibitem [{\citenamefont {te~Velde}\ and\ \citenamefont
  {Baerends}(1991)}]{te_velde_precise_1991}%
  \BibitemOpen
  \bibfield  {author} {\bibinfo {author} {\bibfnamefont {G.}~\bibnamefont
  {te~Velde}}\ and\ \bibinfo {author} {\bibfnamefont {E.~J.}\ \bibnamefont
  {Baerends}},\ }\href {\doibase 10.1103/PhysRevB.44.7888} {\bibfield
  {journal} {\bibinfo  {journal} {Phys. Rev. B}\ }\textbf {\bibinfo {volume}
  {44}},\ \bibinfo {pages} {7888} (\bibinfo {year} {1991})}\BibitemShut
  {NoStop}%
\bibitem [{\citenamefont {Wiesenekker}\ and\ \citenamefont
  {Baerends}(1991)}]{wiesenekker_quadratic_1991}%
  \BibitemOpen
  \bibfield  {author} {\bibinfo {author} {\bibfnamefont {G.}~\bibnamefont
  {Wiesenekker}}\ and\ \bibinfo {author} {\bibfnamefont {E.~J.}\ \bibnamefont
  {Baerends}},\ }\href {\doibase 10.1088/0953-8984/3/35/005} {\bibfield
  {journal} {\bibinfo  {journal} {J. Phys.: Condens. Matter}\ }\textbf
  {\bibinfo {volume} {3}},\ \bibinfo {pages} {6721} (\bibinfo {year}
  {1991})}\BibitemShut {NoStop}%
\bibitem [{\citenamefont {Franchini}\ \emph {et~al.}(2013)\citenamefont
  {Franchini}, \citenamefont {Philipsen},\ and\ \citenamefont
  {Visscher}}]{franchini_becke_2013}%
  \BibitemOpen
  \bibfield  {author} {\bibinfo {author} {\bibfnamefont {M.}~\bibnamefont
  {Franchini}}, \bibinfo {author} {\bibfnamefont {P.~H.~T.}\ \bibnamefont
  {Philipsen}}, \ and\ \bibinfo {author} {\bibfnamefont {L.}~\bibnamefont
  {Visscher}},\ }\href {\doibase 10.1002/jcc.23323} {\bibfield  {journal}
  {\bibinfo  {journal} {Journal of Computational Chemistry}\ }\textbf {\bibinfo
  {volume} {34}},\ \bibinfo {pages} {1819} (\bibinfo {year}
  {2013})}\BibitemShut {NoStop}%
\bibitem [{\citenamefont {Franchini}\ \emph {et~al.}(2014)\citenamefont
  {Franchini}, \citenamefont {Philipsen}, \citenamefont {van Lenthe},\ and\
  \citenamefont {Visscher}}]{franchini_accurate_2014}%
  \BibitemOpen
  \bibfield  {author} {\bibinfo {author} {\bibfnamefont {M.}~\bibnamefont
  {Franchini}}, \bibinfo {author} {\bibfnamefont {P.~H.~T.}\ \bibnamefont
  {Philipsen}}, \bibinfo {author} {\bibfnamefont {E.}~\bibnamefont {van
  Lenthe}}, \ and\ \bibinfo {author} {\bibfnamefont {L.}~\bibnamefont
  {Visscher}},\ }\href {\doibase 10.1021/ct500172n} {\bibfield  {journal}
  {\bibinfo  {journal} {J. Chem. Theory Comput.}\ }\textbf {\bibinfo {volume}
  {10}},\ \bibinfo {pages} {1994} (\bibinfo {year} {2014})}\BibitemShut
  {NoStop}%
\bibitem [{\citenamefont {Paier}\ \emph
  {et~al.}(2006{\natexlab{a}})\citenamefont {Paier}, \citenamefont {Marsman},
  \citenamefont {Hummer}, \citenamefont {Kresse}, \citenamefont {Gerber},\ and\
  \citenamefont {\'{A}ngy\'{a}n}}]{paier_screened_2006}%
  \BibitemOpen
  \bibfield  {author} {\bibinfo {author} {\bibfnamefont {J.}~\bibnamefont
  {Paier}}, \bibinfo {author} {\bibfnamefont {M.}~\bibnamefont {Marsman}},
  \bibinfo {author} {\bibfnamefont {K.}~\bibnamefont {Hummer}}, \bibinfo
  {author} {\bibfnamefont {G.}~\bibnamefont {Kresse}}, \bibinfo {author}
  {\bibfnamefont {I.~C.}\ \bibnamefont {Gerber}}, \ and\ \bibinfo {author}
  {\bibfnamefont {J.~G.}\ \bibnamefont {\'{A}ngy\'{a}n}},\ }\href {\doibase
  10.1063/1.2187006} {\bibfield  {journal} {\bibinfo  {journal} {J. Chem.
  Phys.}\ }\textbf {\bibinfo {volume} {124}},\ \bibinfo {pages} {154709}
  (\bibinfo {year} {2006}{\natexlab{a}})}\BibitemShut {NoStop}%
\bibitem [{\citenamefont {Heyd}\ \emph {et~al.}(2005)\citenamefont {Heyd},
  \citenamefont {Peralta}, \citenamefont {Scuseria},\ and\ \citenamefont
  {Martin}}]{heyd_energy_2005}%
  \BibitemOpen
  \bibfield  {author} {\bibinfo {author} {\bibfnamefont {J.}~\bibnamefont
  {Heyd}}, \bibinfo {author} {\bibfnamefont {J.~E.}\ \bibnamefont {Peralta}},
  \bibinfo {author} {\bibfnamefont {G.~E.}\ \bibnamefont {Scuseria}}, \ and\
  \bibinfo {author} {\bibfnamefont {R.~L.}\ \bibnamefont {Martin}},\ }\href
  {\doibase 10.1063/1.2085170} {\bibfield  {journal} {\bibinfo  {journal} {J.
  Chem. Phys.}\ }\textbf {\bibinfo {volume} {123}},\ \bibinfo {pages} {174101}
  (\bibinfo {year} {2005})}\BibitemShut {NoStop}%
\bibitem [{\citenamefont {Heyd}\ \emph {et~al.}(2006)\citenamefont {Heyd},
  \citenamefont {Scuseria},\ and\ \citenamefont
  {Ernzerhof}}]{heyd_erratum:_2006}%
  \BibitemOpen
  \bibfield  {author} {\bibinfo {author} {\bibfnamefont {J.}~\bibnamefont
  {Heyd}}, \bibinfo {author} {\bibfnamefont {G.~E.}\ \bibnamefont {Scuseria}},
  \ and\ \bibinfo {author} {\bibfnamefont {M.}~\bibnamefont {Ernzerhof}},\
  }\href {\doibase doi:10.1063/1.2204597} {\bibfield  {journal} {\bibinfo
  {journal} {J. Chem. Phys.}\ }\textbf {\bibinfo {volume} {124}},\ \bibinfo
  {pages} {219906} (\bibinfo {year} {2006})}\BibitemShut {NoStop}%
\bibitem [{\citenamefont {Paier}\ \emph
  {et~al.}(2006{\natexlab{b}})\citenamefont {Paier}, \citenamefont {Marsman},
  \citenamefont {Hummer}, \citenamefont {Kresse}, \citenamefont {Gerber},\ and\
  \citenamefont {\'{A}ngy\'{a}n}}]{paier_erratum:_2006}%
  \BibitemOpen
  \bibfield  {author} {\bibinfo {author} {\bibfnamefont {J.}~\bibnamefont
  {Paier}}, \bibinfo {author} {\bibfnamefont {M.}~\bibnamefont {Marsman}},
  \bibinfo {author} {\bibfnamefont {K.}~\bibnamefont {Hummer}}, \bibinfo
  {author} {\bibfnamefont {G.}~\bibnamefont {Kresse}}, \bibinfo {author}
  {\bibfnamefont {I.~C.}\ \bibnamefont {Gerber}}, \ and\ \bibinfo {author}
  {\bibfnamefont {J.~G.}\ \bibnamefont {\'{A}ngy\'{a}n}},\ }\href {\doibase
  doi:10.1063/1.2403866} {\bibfield  {journal} {\bibinfo  {journal} {J. Chem.
  Phys.}\ }\textbf {\bibinfo {volume} {125}},\ \bibinfo {pages} {249901}
  (\bibinfo {year} {2006}{\natexlab{b}})}\BibitemShut {NoStop}%
\bibitem [{\citenamefont {Madsen}\ and\ \citenamefont
  {Singh}(2006)}]{Madsen200667}%
  \BibitemOpen
  \bibfield  {author} {\bibinfo {author} {\bibfnamefont {G.~K.}\ \bibnamefont
  {Madsen}}\ and\ \bibinfo {author} {\bibfnamefont {D.~J.}\ \bibnamefont
  {Singh}},\ }\href {\doibase http://dx.doi.org/10.1016/j.cpc.2006.03.007}
  {\bibfield  {journal} {\bibinfo  {journal} {Comput. Phys. Commun.}\ }\textbf
  {\bibinfo {volume} {175}},\ \bibinfo {pages} {67 } (\bibinfo {year}
  {2006})}\BibitemShut {NoStop}%
\bibitem [{\citenamefont {Singh}(2010)}]{singh_doping-dependent_2010}%
  \BibitemOpen
  \bibfield  {author} {\bibinfo {author} {\bibfnamefont {D.~J.}\ \bibnamefont
  {Singh}},\ }\href {\doibase 10.1103/PhysRevB.81.195217} {\bibfield  {journal}
  {\bibinfo  {journal} {Phys. Rev. B}\ }\textbf {\bibinfo {volume} {81}},\
  \bibinfo {pages} {195217} (\bibinfo {year} {2010})}\BibitemShut {NoStop}%
\bibitem [{\citenamefont {Chen}\ \emph {et~al.}(2016)\citenamefont {Chen},
  \citenamefont {Pöhls}, \citenamefont {Hautier}, \citenamefont {Broberg},
  \citenamefont {Bajaj}, \citenamefont {Aydemir}, \citenamefont {Gibbs},
  \citenamefont {Zhu}, \citenamefont {Asta}, \citenamefont {Snyder},
  \citenamefont {Meredig}, \citenamefont {White}, \citenamefont {Persson},\
  and\ \citenamefont {Jain}}]{chen_understanding_2016}%
  \BibitemOpen
  \bibfield  {author} {\bibinfo {author} {\bibfnamefont {W.}~\bibnamefont
  {Chen}}, \bibinfo {author} {\bibfnamefont {J.-H.}\ \bibnamefont {Pöhls}},
  \bibinfo {author} {\bibfnamefont {G.}~\bibnamefont {Hautier}}, \bibinfo
  {author} {\bibfnamefont {D.}~\bibnamefont {Broberg}}, \bibinfo {author}
  {\bibfnamefont {S.}~\bibnamefont {Bajaj}}, \bibinfo {author} {\bibfnamefont
  {U.}~\bibnamefont {Aydemir}}, \bibinfo {author} {\bibfnamefont {Z.~M.}\
  \bibnamefont {Gibbs}}, \bibinfo {author} {\bibfnamefont {H.}~\bibnamefont
  {Zhu}}, \bibinfo {author} {\bibfnamefont {M.}~\bibnamefont {Asta}}, \bibinfo
  {author} {\bibfnamefont {G.~J.}\ \bibnamefont {Snyder}}, \bibinfo {author}
  {\bibfnamefont {B.}~\bibnamefont {Meredig}}, \bibinfo {author} {\bibfnamefont
  {M.~A.}\ \bibnamefont {White}}, \bibinfo {author} {\bibfnamefont
  {K.}~\bibnamefont {Persson}}, \ and\ \bibinfo {author} {\bibfnamefont
  {A.}~\bibnamefont {Jain}},\ }\href {\doibase 10.1039/C5TC04339E} {\bibfield
  {journal} {\bibinfo  {journal} {J. Mater. Chem. C}\ }\textbf {\bibinfo
  {volume} {4}},\ \bibinfo {pages} {4414} (\bibinfo {year} {2016})}\BibitemShut
  {NoStop}%
\bibitem [{\citenamefont {Bilc}\ \emph {et~al.}(2015)\citenamefont {Bilc},
  \citenamefont {Hautier}, \citenamefont {Waroquiers}, \citenamefont
  {Rignanese},\ and\ \citenamefont {Ghosez}}]{bilc_low-dimensional_2015}%
  \BibitemOpen
  \bibfield  {author} {\bibinfo {author} {\bibfnamefont {D.~I.}\ \bibnamefont
  {Bilc}}, \bibinfo {author} {\bibfnamefont {G.}~\bibnamefont {Hautier}},
  \bibinfo {author} {\bibfnamefont {D.}~\bibnamefont {Waroquiers}}, \bibinfo
  {author} {\bibfnamefont {G.-M.}\ \bibnamefont {Rignanese}}, \ and\ \bibinfo
  {author} {\bibfnamefont {P.}~\bibnamefont {Ghosez}},\ }\href {\doibase
  10.1103/PhysRevLett.114.136601} {\bibfield  {journal} {\bibinfo  {journal}
  {Phys. Rev. Lett.}\ }\textbf {\bibinfo {volume} {114}},\ \bibinfo {pages}
  {136601} (\bibinfo {year} {2015})}\BibitemShut {NoStop}%
\bibitem [{\citenamefont {Togo}\ and\ \citenamefont {Tanaka}(2015)}]{phonopy}%
  \BibitemOpen
  \bibfield  {author} {\bibinfo {author} {\bibfnamefont {A.}~\bibnamefont
  {Togo}}\ and\ \bibinfo {author} {\bibfnamefont {I.}~\bibnamefont {Tanaka}},\
  }\href {\doibase https://doi.org/10.1016/j.scriptamat.2015.07.021} {\bibfield
   {journal} {\bibinfo  {journal} {Scripta Materialia}\ }\textbf {\bibinfo
  {volume} {108}},\ \bibinfo {pages} {1 } (\bibinfo {year} {2015})}\BibitemShut
  {NoStop}%
\bibitem [{\citenamefont {Zhou}\ \emph {et~al.}(2014)\citenamefont {Zhou},
  \citenamefont {Nielson}, \citenamefont {Xia},\ and\ \citenamefont
  {Ozoli\ifmmode \mbox{\c{n}}\else \c{n}\fi{}\ifmmode~\check{s}\else
  \v{s}\fi{}}}]{PhysRevLett.113.185501}%
  \BibitemOpen
  \bibfield  {author} {\bibinfo {author} {\bibfnamefont {F.}~\bibnamefont
  {Zhou}}, \bibinfo {author} {\bibfnamefont {W.}~\bibnamefont {Nielson}},
  \bibinfo {author} {\bibfnamefont {Y.}~\bibnamefont {Xia}}, \ and\ \bibinfo
  {author} {\bibfnamefont {V.}~\bibnamefont {Ozoli\ifmmode \mbox{\c{n}}\else
  \c{n}\fi{}\ifmmode~\check{s}\else \v{s}\fi{}}},\ }\href {\doibase
  10.1103/PhysRevLett.113.185501} {\bibfield  {journal} {\bibinfo  {journal}
  {Phys. Rev. Lett.}\ }\textbf {\bibinfo {volume} {113}},\ \bibinfo {pages}
  {185501} (\bibinfo {year} {2014})}\BibitemShut {NoStop}%
\bibitem [{\citenamefont {Li}\ \emph {et~al.}(2014)\citenamefont {Li},
  \citenamefont {Carrete}, \citenamefont {Katcho},\ and\ \citenamefont
  {Mingo}}]{ShengBTE_2014}%
  \BibitemOpen
  \bibfield  {author} {\bibinfo {author} {\bibfnamefont {W.}~\bibnamefont
  {Li}}, \bibinfo {author} {\bibfnamefont {J.}~\bibnamefont {Carrete}},
  \bibinfo {author} {\bibfnamefont {N.~A.}\ \bibnamefont {Katcho}}, \ and\
  \bibinfo {author} {\bibfnamefont {N.}~\bibnamefont {Mingo}},\ }\href
  {\doibase 10.1016/j.cpc.2014.02.015} {\bibfield  {journal} {\bibinfo
  {journal} {Comp. Phys. Commun.}\ }\textbf {\bibinfo {volume} {185}},\
  \bibinfo {pages} {1747} (\bibinfo {year} {2014})}\BibitemShut {NoStop}%
\bibitem [{\citenamefont {Hurng}\ and\ \citenamefont
  {Corbett}(1989)}]{hurng_alkaline-earth-metal_1989}%
  \BibitemOpen
  \bibfield  {author} {\bibinfo {author} {\bibfnamefont {W.~M.}\ \bibnamefont
  {Hurng}}\ and\ \bibinfo {author} {\bibfnamefont {J.~D.}\ \bibnamefont
  {Corbett}},\ }\href@noop {} {\bibfield  {journal} {\bibinfo  {journal} {Chem.
  Mater.}\ }\textbf {\bibinfo {volume} {1}},\ \bibinfo {pages} {311} (\bibinfo
  {year} {1989})}\BibitemShut {NoStop}%
\bibitem [{\citenamefont {Hadenfeldt}\ and\ \citenamefont
  {Vollert}(1988)}]{hadenfeldt_darstellung_1988}%
  \BibitemOpen
  \bibfield  {author} {\bibinfo {author} {\bibfnamefont {C.}~\bibnamefont
  {Hadenfeldt}}\ and\ \bibinfo {author} {\bibfnamefont {H.~O.}\ \bibnamefont
  {Vollert}},\ }\href {\doibase 10.1016/0022-5088(88)90126-9} {\bibfield
  {journal} {\bibinfo  {journal} {Journal of the Less Common Metals}\ }\textbf
  {\bibinfo {volume} {144}},\ \bibinfo {pages} {143} (\bibinfo {year}
  {1988})}\BibitemShut {NoStop}%
\bibitem [{\citenamefont {Xia}\ and\ \citenamefont
  {Bobev}(2007)}]{xia_existence_2007}%
  \BibitemOpen
  \bibfield  {author} {\bibinfo {author} {\bibfnamefont {S.}~\bibnamefont
  {Xia}}\ and\ \bibinfo {author} {\bibfnamefont {S.}~\bibnamefont {Bobev}},\
  }\href {\doibase 10.1016/j.jallcom.2006.02.046} {\bibfield  {journal}
  {\bibinfo  {journal} {Journal of Alloys and Compounds}\ }\textbf {\bibinfo
  {volume} {427}},\ \bibinfo {pages} {67} (\bibinfo {year} {2007})}\BibitemShut
  {NoStop}%
\bibitem [{\citenamefont {Hadenfeldt}\ and\ \citenamefont
  {Tersch\"{u}ren}(1991)}]{hadenfeldt_darstellung_1991}%
  \BibitemOpen
  \bibfield  {author} {\bibinfo {author} {\bibfnamefont {C.}~\bibnamefont
  {Hadenfeldt}}\ and\ \bibinfo {author} {\bibfnamefont {H.-U.}\ \bibnamefont
  {Tersch\"{u}ren}},\ }\href {\doibase 10.1002/zaac.19915970110} {\bibfield
  {journal} {\bibinfo  {journal} {Zeitschrift f\"{u}r anorganische und
  allgemeine Chemie}\ }\textbf {\bibinfo {volume} {597}},\ \bibinfo {pages}
  {69} (\bibinfo {year} {1991})}\BibitemShut {NoStop}%
\bibitem [{\citenamefont {Wied}\ \emph {et~al.}(2014)\citenamefont {Wied},
  \citenamefont {Nuss}, \citenamefont {H\"{o}nle},\ and\ \citenamefont
  {Schnering}}]{wied_crystal_2014}%
  \BibitemOpen
  \bibfield  {author} {\bibinfo {author} {\bibfnamefont {M.}~\bibnamefont
  {Wied}}, \bibinfo {author} {\bibfnamefont {J.}~\bibnamefont {Nuss}}, \bibinfo
  {author} {\bibfnamefont {W.}~\bibnamefont {H\"{o}nle}}, \ and\ \bibinfo
  {author} {\bibfnamefont {H.~G.~v.}\ \bibnamefont {Schnering}},\ }\href
  {\doibase 10.1524/ncrs.2011.0195} {\bibfield  {journal} {\bibinfo  {journal}
  {Zeitschrift f\"{u}r Kristallographie - New Crystal Structures}\ }\textbf
  {\bibinfo {volume} {226}},\ \bibinfo {pages} {437} (\bibinfo {year}
  {2014})}\BibitemShut {NoStop}%
\bibitem [{\citenamefont {Klos}(2018)}]{klos_ternare_nodate}%
  \BibitemOpen
  \bibfield  {author} {\bibinfo {author} {\bibfnamefont {S.}~\bibnamefont
  {Klos}},\ }\href@noop {} {\enquote {\bibinfo {title} {Tern\"{a}re
  {Zintl}-{Phasen} ({Erd}){Alkalimetall}-{Triel}-{Pentel} und deren partielle
  {Oxidation} zu {Pentelidgallaten}},}\ } (\bibinfo {year} {2018}),\ \bibinfo
  {note} {{PhD Thesis}}\BibitemShut {NoStop}%
\bibitem [{\citenamefont {R\"{o}hr}\ and\ \citenamefont
  {George}(2010)}]{rohr_crystal_2010}%
  \BibitemOpen
  \bibfield  {author} {\bibinfo {author} {\bibfnamefont {C.}~\bibnamefont
  {R\"{o}hr}}\ and\ \bibinfo {author} {\bibfnamefont {R.}~\bibnamefont
  {George}},\ }\href {\doibase 10.1524/zkri.1996.211.7.478} {\bibfield
  {journal} {\bibinfo  {journal} {Zeitschrift für Kristallographie -
  Crystalline Materials}\ }\textbf {\bibinfo {volume} {211}},\ \bibinfo {pages}
  {478} (\bibinfo {year} {2010})}\BibitemShut {NoStop}%
\bibitem [{\citenamefont {Li}\ \emph {et~al.}(2003)\citenamefont {Li},
  \citenamefont {Mudring},\ and\ \citenamefont {Corbett}}]{li_valence_2003}%
  \BibitemOpen
  \bibfield  {author} {\bibinfo {author} {\bibfnamefont {B.}~\bibnamefont
  {Li}}, \bibinfo {author} {\bibfnamefont {A.-V.}\ \bibnamefont {Mudring}}, \
  and\ \bibinfo {author} {\bibfnamefont {J.~D.}\ \bibnamefont {Corbett}},\
  }\href {\doibase 10.1021/ic0301472} {\bibfield  {journal} {\bibinfo
  {journal} {Inorg. Chem.}\ }\textbf {\bibinfo {volume} {42}},\ \bibinfo
  {pages} {6940} (\bibinfo {year} {2003})}\BibitemShut {NoStop}%
\bibitem [{\citenamefont {Janiak}\ \emph {et~al.}(2012)\citenamefont {Janiak},
  \citenamefont {Meyer}, \citenamefont {Gudat}, \citenamefont {Alsfasser},\
  and\ \citenamefont {Meyer}}]{janiak_riedel_2012}%
  \BibitemOpen
  \bibfield  {author} {\bibinfo {author} {\bibfnamefont {C.}~\bibnamefont
  {Janiak}}, \bibinfo {author} {\bibfnamefont {H.-J.}\ \bibnamefont {Meyer}},
  \bibinfo {author} {\bibfnamefont {D.}~\bibnamefont {Gudat}}, \bibinfo
  {author} {\bibfnamefont {R.}~\bibnamefont {Alsfasser}}, \ and\ \bibinfo
  {author} {\bibfnamefont {H.-J.}\ \bibnamefont {Meyer}},\ }\href {\doibase
  10.1515/9783110249019} {\emph {\bibinfo {title} {Riedel {Moderne}
  {Anorganische} {Chemie}}}}\ (\bibinfo  {publisher} {De Gruyter},\ \bibinfo
  {address} {Berlin, Boston},\ \bibinfo {year} {2012})\BibitemShut {NoStop}%
\bibitem [{AM1()}]{AM1.5}%
  \BibitemOpen
  \href@noop {} {}\bibinfo {note} {ASTM Standard G173-03(2008), Standard Tables
  for Reference Solar Spectral Irradiances: Direct Normal and Hemispherical on
  37$^{\circ}$ Tilted Surface (ASTM International, West Conshohocken, PA,
  2008).}\BibitemShut {Stop}%
\bibitem [{\citenamefont {Hulliger}(1979)}]{hulliger_new_1979}%
  \BibitemOpen
  \bibfield  {author} {\bibinfo {author} {\bibfnamefont {F.}~\bibnamefont
  {Hulliger}},\ }\href {\doibase 10.1016/0025-5408(79)90127-2} {\bibfield
  {journal} {\bibinfo  {journal} {Materials Research Bulletin}\ }\textbf
  {\bibinfo {volume} {14}},\ \bibinfo {pages} {259} (\bibinfo {year}
  {1979})}\BibitemShut {NoStop}%
\bibitem [{\citenamefont {Garza}\ and\ \citenamefont
  {Scuseria}(2016)}]{garza_predicting_2016}%
  \BibitemOpen
  \bibfield  {author} {\bibinfo {author} {\bibfnamefont {A.~J.}\ \bibnamefont
  {Garza}}\ and\ \bibinfo {author} {\bibfnamefont {G.~E.}\ \bibnamefont
  {Scuseria}},\ }\href {\doibase 10.1021/acs.jpclett.6b01807} {\bibfield
  {journal} {\bibinfo  {journal} {J. Phys. Chem. Lett.}\ }\textbf {\bibinfo
  {volume} {7}},\ \bibinfo {pages} {4165} (\bibinfo {year} {2016})}\BibitemShut
  {NoStop}%
\bibitem [{\citenamefont {Kauzlarich}\ \emph {et~al.}(2007)\citenamefont
  {Kauzlarich}, \citenamefont {Brown},\ and\ \citenamefont
  {Snyder}}]{kauzlarich_zintl_2007}%
  \BibitemOpen
  \bibfield  {author} {\bibinfo {author} {\bibfnamefont {S.~M.}\ \bibnamefont
  {Kauzlarich}}, \bibinfo {author} {\bibfnamefont {S.~R.}\ \bibnamefont
  {Brown}}, \ and\ \bibinfo {author} {\bibfnamefont {G.~J.}\ \bibnamefont
  {Snyder}},\ }\href {\doibase 10.1039/B702266B} {\bibfield  {journal}
  {\bibinfo  {journal} {Dalton Trans.}\ }\textbf {\bibinfo {volume} {0}},\
  \bibinfo {pages} {2099} (\bibinfo {year} {2007})}\BibitemShut {NoStop}%
\bibitem [{\citenamefont {Snyder}\ and\ \citenamefont
  {Toberer}(2008)}]{snyder_complex_2008}%
  \BibitemOpen
  \bibfield  {author} {\bibinfo {author} {\bibfnamefont {G.~J.}\ \bibnamefont
  {Snyder}}\ and\ \bibinfo {author} {\bibfnamefont {E.~S.}\ \bibnamefont
  {Toberer}},\ }\href {\doibase 10.1038/nmat2090} {\bibfield  {journal}
  {\bibinfo  {journal} {Nat Mater}\ }\textbf {\bibinfo {volume} {7}},\ \bibinfo
  {pages} {105} (\bibinfo {year} {2008})}\BibitemShut {NoStop}%
\bibitem [{\citenamefont {Rowe}(2005)}]{rowe2005thermoelectrics}%
  \BibitemOpen
  \bibfield  {author} {\bibinfo {author} {\bibfnamefont {D.}~\bibnamefont
  {Rowe}},\ }\href {http://books.google.gr/books?id=0iwERQe5IKQC} {\emph
  {\bibinfo {title} {Thermoelectrics Handbook: Macro to Nano}}}\ (\bibinfo
  {publisher} {CRC Press},\ \bibinfo {year} {2005})\BibitemShut {NoStop}%
\bibitem [{\citenamefont {Zhao}\ \emph {et~al.}(2013)\citenamefont {Zhao},
  \citenamefont {Hao}, \citenamefont {Lo}, \citenamefont {Wu}, \citenamefont
  {Zhou}, \citenamefont {Lee}, \citenamefont {Li}, \citenamefont {Biswas},
  \citenamefont {Hogan}, \citenamefont {Uher}, \citenamefont {Wolverton},
  \citenamefont {Dravid},\ and\ \citenamefont {Kanatzidis}}]{zhao_high_2013}%
  \BibitemOpen
  \bibfield  {author} {\bibinfo {author} {\bibfnamefont {L.-D.}\ \bibnamefont
  {Zhao}}, \bibinfo {author} {\bibfnamefont {S.}~\bibnamefont {Hao}}, \bibinfo
  {author} {\bibfnamefont {S.-H.}\ \bibnamefont {Lo}}, \bibinfo {author}
  {\bibfnamefont {C.-I.}\ \bibnamefont {Wu}}, \bibinfo {author} {\bibfnamefont
  {X.}~\bibnamefont {Zhou}}, \bibinfo {author} {\bibfnamefont {Y.}~\bibnamefont
  {Lee}}, \bibinfo {author} {\bibfnamefont {H.}~\bibnamefont {Li}}, \bibinfo
  {author} {\bibfnamefont {K.}~\bibnamefont {Biswas}}, \bibinfo {author}
  {\bibfnamefont {T.~P.}\ \bibnamefont {Hogan}}, \bibinfo {author}
  {\bibfnamefont {C.}~\bibnamefont {Uher}}, \bibinfo {author} {\bibfnamefont
  {C.}~\bibnamefont {Wolverton}}, \bibinfo {author} {\bibfnamefont {V.~P.}\
  \bibnamefont {Dravid}}, \ and\ \bibinfo {author} {\bibfnamefont {M.~G.}\
  \bibnamefont {Kanatzidis}},\ }\href {\doibase 10.1021/ja403134b} {\bibfield
  {journal} {\bibinfo  {journal} {J. Am. Chem. Soc.}\ }\textbf {\bibinfo
  {volume} {135}},\ \bibinfo {pages} {7364} (\bibinfo {year}
  {2013})}\BibitemShut {NoStop}%
\bibitem [{\citenamefont {Yin}\ \emph {et~al.}(2016)\citenamefont {Yin},
  \citenamefont {Su}, \citenamefont {Yan}, \citenamefont {You}, \citenamefont
  {Zhang}, \citenamefont {Uher}, \citenamefont {Kanatzidis},\ and\
  \citenamefont {Tang}}]{yin_optimization_2016}%
  \BibitemOpen
  \bibfield  {author} {\bibinfo {author} {\bibfnamefont {K.}~\bibnamefont
  {Yin}}, \bibinfo {author} {\bibfnamefont {X.}~\bibnamefont {Su}}, \bibinfo
  {author} {\bibfnamefont {Y.}~\bibnamefont {Yan}}, \bibinfo {author}
  {\bibfnamefont {Y.}~\bibnamefont {You}}, \bibinfo {author} {\bibfnamefont
  {Q.}~\bibnamefont {Zhang}}, \bibinfo {author} {\bibfnamefont
  {C.}~\bibnamefont {Uher}}, \bibinfo {author} {\bibfnamefont {M.~G.}\
  \bibnamefont {Kanatzidis}}, \ and\ \bibinfo {author} {\bibfnamefont
  {X.}~\bibnamefont {Tang}},\ }\href {\doibase 10.1021/acs.chemmater.6b02308}
  {\bibfield  {journal} {\bibinfo  {journal} {Chem. Mater.}\ }\textbf {\bibinfo
  {volume} {28}},\ \bibinfo {pages} {5538} (\bibinfo {year}
  {2016})}\BibitemShut {NoStop}%
\bibitem [{\citenamefont {Rao}\ \emph {et~al.}(2006)\citenamefont {Rao},
  \citenamefont {Ji},\ and\ \citenamefont {Tritt}}]{rao_properties_2006}%
  \BibitemOpen
  \bibfield  {author} {\bibinfo {author} {\bibfnamefont {A.~M.}\ \bibnamefont
  {Rao}}, \bibinfo {author} {\bibfnamefont {X.}~\bibnamefont {Ji}}, \ and\
  \bibinfo {author} {\bibfnamefont {T.~M.}\ \bibnamefont {Tritt}},\ }\href
  {\doibase 10.1557/mrs2006.48} {\bibfield  {journal} {\bibinfo  {journal}
  {{MRS} Bulletin}\ }\textbf {\bibinfo {volume} {31}},\ \bibinfo {pages} {218}
  (\bibinfo {year} {2006})}\BibitemShut {NoStop}%
\bibitem [{\citenamefont {Toberer}\ \emph {et~al.}(2011)\citenamefont
  {Toberer}, \citenamefont {Zevalkink},\ and\ \citenamefont
  {Snyder}}]{toberer_phonon_2011}%
  \BibitemOpen
  \bibfield  {author} {\bibinfo {author} {\bibfnamefont {E.~S.}\ \bibnamefont
  {Toberer}}, \bibinfo {author} {\bibfnamefont {A.}~\bibnamefont {Zevalkink}},
  \ and\ \bibinfo {author} {\bibfnamefont {G.~J.}\ \bibnamefont {Snyder}},\
  }\href {\doibase 10.1039/C1JM11754H} {\bibfield  {journal} {\bibinfo
  {journal} {J. Mater. Chem.}\ }\textbf {\bibinfo {volume} {21}},\ \bibinfo
  {pages} {15843} (\bibinfo {year} {2011})}\BibitemShut {NoStop}%
\bibitem [{\citenamefont {Toberer}\ \emph {et~al.}(2010)\citenamefont
  {Toberer}, \citenamefont {May},\ and\ \citenamefont
  {Snyder}}]{toberer_zintl_2010}%
  \BibitemOpen
  \bibfield  {author} {\bibinfo {author} {\bibfnamefont {E.~S.}\ \bibnamefont
  {Toberer}}, \bibinfo {author} {\bibfnamefont {A.~F.}\ \bibnamefont {May}}, \
  and\ \bibinfo {author} {\bibfnamefont {G.~J.}\ \bibnamefont {Snyder}},\
  }\href {\doibase 10.1021/cm901956r} {\bibfield  {journal} {\bibinfo
  {journal} {Chem. Mater.}\ }\textbf {\bibinfo {volume} {22}},\ \bibinfo
  {pages} {624} (\bibinfo {year} {2010})}\BibitemShut {NoStop}%
\bibitem [{\citenamefont {Zevalkink}\ \emph {et~al.}(2014)\citenamefont
  {Zevalkink}, \citenamefont {Swallow}, \citenamefont {Ohno}, \citenamefont
  {Aydemir}, \citenamefont {Bux},\ and\ \citenamefont
  {Snyder}}]{zevalkink_thermoelectric_2014}%
  \BibitemOpen
  \bibfield  {author} {\bibinfo {author} {\bibfnamefont {A.}~\bibnamefont
  {Zevalkink}}, \bibinfo {author} {\bibfnamefont {J.}~\bibnamefont {Swallow}},
  \bibinfo {author} {\bibfnamefont {S.}~\bibnamefont {Ohno}}, \bibinfo {author}
  {\bibfnamefont {U.}~\bibnamefont {Aydemir}}, \bibinfo {author} {\bibfnamefont
  {S.}~\bibnamefont {Bux}}, \ and\ \bibinfo {author} {\bibfnamefont {G.~J.}\
  \bibnamefont {Snyder}},\ }\href {\doibase 10.1039/C4DT02206H} {\bibfield
  {journal} {\bibinfo  {journal} {Dalton Trans.}\ }\textbf {\bibinfo {volume}
  {43}},\ \bibinfo {pages} {15872} (\bibinfo {year} {2014})}\BibitemShut
  {NoStop}%
\bibitem [{\citenamefont {Bobev}\ \emph {et~al.}(2004)\citenamefont {Bobev},
  \citenamefont {Thompson}, \citenamefont {Sarrao}, \citenamefont {Olmstead},
  \citenamefont {Hope},\ and\ \citenamefont {Kauzlarich}}]{bobev_probing_2004}%
  \BibitemOpen
  \bibfield  {author} {\bibinfo {author} {\bibfnamefont {S.}~\bibnamefont
  {Bobev}}, \bibinfo {author} {\bibfnamefont {J.~D.}\ \bibnamefont {Thompson}},
  \bibinfo {author} {\bibfnamefont {J.~L.}\ \bibnamefont {Sarrao}}, \bibinfo
  {author} {\bibfnamefont {M.~M.}\ \bibnamefont {Olmstead}}, \bibinfo {author}
  {\bibfnamefont {H.}~\bibnamefont {Hope}}, \ and\ \bibinfo {author}
  {\bibfnamefont {S.~M.}\ \bibnamefont {Kauzlarich}},\ }\href {\doibase
  10.1021/ic049836j} {\bibfield  {journal} {\bibinfo  {journal} {Inorg. Chem.}\
  }\textbf {\bibinfo {volume} {43}},\ \bibinfo {pages} {5044} (\bibinfo {year}
  {2004})}\BibitemShut {NoStop}%
\bibitem [{\citenamefont {Ohno}\ \emph {et~al.}(2017)\citenamefont {Ohno},
  \citenamefont {Aydemir}, \citenamefont {Amsler}, \citenamefont {P\"{o}hls},
  \citenamefont {Chanakian}, \citenamefont {Zevalkink}, \citenamefont {White},
  \citenamefont {Bux}, \citenamefont {Wolverton},\ and\ \citenamefont
  {Snyder}}]{ohno_achieving_2017}%
  \BibitemOpen
  \bibfield  {author} {\bibinfo {author} {\bibfnamefont {S.}~\bibnamefont
  {Ohno}}, \bibinfo {author} {\bibfnamefont {U.}~\bibnamefont {Aydemir}},
  \bibinfo {author} {\bibfnamefont {M.}~\bibnamefont {Amsler}}, \bibinfo
  {author} {\bibfnamefont {J.-H.}\ \bibnamefont {P\"{o}hls}}, \bibinfo {author}
  {\bibfnamefont {S.}~\bibnamefont {Chanakian}}, \bibinfo {author}
  {\bibfnamefont {A.}~\bibnamefont {Zevalkink}}, \bibinfo {author}
  {\bibfnamefont {M.~A.}\ \bibnamefont {White}}, \bibinfo {author}
  {\bibfnamefont {S.~K.}\ \bibnamefont {Bux}}, \bibinfo {author} {\bibfnamefont
  {C.}~\bibnamefont {Wolverton}}, \ and\ \bibinfo {author} {\bibfnamefont
  {G.~J.}\ \bibnamefont {Snyder}},\ }\href {\doibase 10.1002/adfm.201606361}
  {\bibfield  {journal} {\bibinfo  {journal} {Adv. Funct. Mater.}\ }\textbf
  {\bibinfo {volume} {27}},\ \bibinfo {pages} {1606361} (\bibinfo {year}
  {2017})}\BibitemShut {NoStop}%
\end{thebibliography}
%

\end{document}